\documentclass[structabstract]{aa}  
\usepackage{bigstrut}
\usepackage{graphicx}
\usepackage{txfonts}
\usepackage[round]{natbib}
\usepackage{url}
\usepackage{textcomp}
\begin{document}
   \title{Exploring wind-driving dust species in cool luminous giants}

   \subtitle{III. Wind models for M-type AGB stars: dynamic and photometric properties}

   \author{S. Bladh
          \inst{1}
          \and
          S. H\"ofner\inst{1}
          \and
          B. Aringer\inst{2}
          \and
          K. Eriksson\inst{1}
          }
   \institute{Department of Physics and Astronomy, Division of Astronomy and Space Physics, Uppsala University,
              Box 516, SE-75120, Uppsala, Sweden\\
              \email{sara.bladh@physics.uu.se}
\and
             University of Vienna, Department of Astrophysics, T\"urkenschanzstra{\ss}e 17, A-1180 Wien, Austria
             }

   \date{Received September 4, 2014; accepted XXXX, 2014}

% \abstract{}{}{}{}{} 
% 5 {} token are mandatory
 
\titlerunning{Exploring wind-driving dust species in cool luminous giants III.}
\authorrunning{S. Bladh et al. } 
 
  \abstract
   {Stellar winds observed in asymptotic giant branch (AGB) stars are usually attributed to a combination of stellar pulsations and radiation pressure on dust. Shock waves triggered by pulsations propagate through the atmosphere, compressing the gas and lifting it to cooler regions which creates favourable conditions for grain growth. If sufficient radiative acceleration is exerted on the newly formed grains through absorption or scattering of stellar photons, an outflow can be triggered. Strong candidates for wind-driving dust species in M-type AGB stars are magnesium silicates (Mg$_2$SiO$_4$ and MgSiO$_3$). Such grains can form close to the stellar surface, they consist of abundant materials and, if they grow to sizes comparable to the wavelength of the stellar flux maximum, they experience strong acceleration by photon scattering.}
   {The purpose of this study is to investigate if photon scattering on Mg$_2$SiO$_4$ grains can produce realistic outflows for a wide range of stellar parameters in M-type AGB stars. }
   {We use a frequency-dependent radiation-hydrodynamics code with a detailed description for the growth of Mg$_2$SiO$_4$ grains to calculate the first extensive set of time-dependent wind models for M-type AGB stars. This set includes 139 solar-mass models, with three different luminosities ($5000\,$L$_{\odot}$, $7000\,$L$_{\odot}$, and $10000\,$L$_{\odot}$) and effective temperatures ranging from 2600\,K to 3200\,K. The resulting wind properties, visual and near-IR photometry and mid-IR spectra are compared with observations.}
   {We show that the models can produce outflows for a wide range of stellar parameters. We also demonstrate that they reproduce observed mass-loss rates and wind velocities, as well as visual and near-IR photometry. However, the current models do not show the characteristic silicate features at 10 and 18 $\mu$m as a result of the cool temperature of Mg$_2$SiO$_4$ grains in the wind. Including a small amount of Fe in the grains further out in the circumstellar envelope will increase the grain temperature and result in pronounced silicate features, without significantly affecting the photometry in the visual and near-IR wavelength regions.}
  % conclusions heading (optional)
   {Outflows driven by photon scattering on Mg$_2$SiO$_4$ grains are a viable wind scenario for M-type AGB stars, given the success of the current models in reproducing observed mass-loss rates, wind velocities, and photometry. Both synthetic and observed photometry suggest that the dusty envelopes of M-type AGB stars are quite transparent at visual and near-IR wavelengths, otherwise the variations in visual flux would not be dominated by molecular features.}


   \maketitle
\section{Introduction}
It is generally assumed that the slow but massive outflows of gas and dust from cool luminous AGB stars are accelerated by radiation pressure on dust grains. The total momentum of the photons emitted by these stars easily matches, or even exceeds, the typical momentum of the stellar winds, and solid particles with the right optical properties can be very efficient in gaining momentum from stellar photons through absorption and scattering processes. 

Observationally, grain materials are usually identified through their characteristic lattice modes in the mid-IR, e.g. the well-known silicate features at about 10 and 18$\,\mu$m which are due to stretching and bending modes in the SiO$_4$ tetrahedron. Mid-IR spectra of AGB stars give valuable insights into the complex dust chemistry in the circumstellar envelopes \citep[see e.g.][for an overview]{dor10,agbgrain}. For determining which grains may contribute to wind acceleration, however, important constraints come from the visual and near-infrared wavelength regions which are crucial for the energy and momentum budget of the atmosphere since the radiative flux maximum of the star is at about $1-2\,\mu$m. 

This paper is the third in a series dedicated to identifying wind-driving dust species in M-type AGB stars by using a combination of different dynamical models and observational data. In the first paper \citep{bladh12} we focused on dynamical constraints for material properties, using a simple parameterised description of the dust component in frequency-dependent radiation-hydrodynamical models for pulsating atmospheres and winds. Since dust temperatures are strongly affected by the wavelength-dependence of the grain opacities (causing greenhouse or inverse greenhouse effects), we found that many dust species cannot condense sufficiently close to the star to trigger or accelerate a wind. In particular, Fe-bearing silicates suffer from a severe greenhouse effect as a result of the steep slope of the absorption coefficient at near-IR wavelengths, moving their condensation zone much further away from the star than for Fe-free magnesium silicates, which was also demonstrated by \cite{woi06fe}. This effect, by definition, does not appear in wind models with grey radiative transfer, leading to a severe underestimation of the condensation distance for Fe-bearing silicates.

In the second paper \citep{bladh13} we presented synthetic spectra and visual and near-IR photometry resulting from these parameterised models (set P) and from more detailed wind models driven by photon scattering on Mg$_2$SiO$_4$ grains (set D), first presented in \cite{hof08bg}. Comparing the synthetic photometry with observations of M-type AGB stars we found that the latter models (set D) give a good agreement. In particular, they can reproduce large variations of the $(V-K)$ colour during the pulsation cycle due to abundance variations of TiO. This is a strong indication that the circumstellar envelopes of M-type AGB stars, and, consequently, the wind-driving dust grains are rather transparent, excluding true absorption by dust as the main source of momentum. 
Results from the parameterised models (set P) support this conclusion, which makes photon scattering on Fe-free silicates a prime candidate for the wind driving mechanism. In order for the grains to be efficient at scattering this scenario requires particles of sizes in the range of $0.1-1\,\mu$m. Recently \cite{norr12} claimed the detection of grains with sizes of $\sim0.3\,\mu$m in the close circumstellar environment of 3 AGB stars, using multi-wavelength aperture-masking polarimetric interferometry.

Given the success in reproducing dynamic and photometric properties for the small sample of wind models driven by photon scattering on Mg$_2$SiO$_4$ grains, we here present the first extensive set of time-dependent wind models for M-type AGB stars based on this mechanism. The aim of this study is to demonstrate that outflows can be produced not only for a few selected cases, but also for models with a wide range of stellar parameters. We also investigate how the observable properties of these models are affected by various parameters, and we compare our results with observations, in particular dynamic properties, visual and near-IR photometry, and mid-IR spectra. 

The paper is organised as follows: in Sect.~\ref{s_par} we introduce the basic physical assumptions and input parameters for the dynamical models of M-type AGB stars. In Secs.~\ref{s_dyn}-\ref{s_mir} we evaluate the wind properties, the visual and near-IR photometry and the mid-IR spectra produced by these models and compare with observed values. In Sect.~\ref{s_gtemp} we explore constraints on the position of the wind acceleration zone and in Sect.~\ref{s_sum} we provide a summary of our conclusions.

\section{Modelling method and parameters}
\label{s_par}
The models presented in this paper are similar to model set D in \cite{bladh13}, corresponding to the models first presented in \cite{hof08bg}.  We only give a short summary of their main ingredients here. For more detailed information concerning the dynamical models and the \textit{a posteriori} radiative transfer, see previous articles \citep{hof03,hof08bg,bladh12,bladh13} and references therein.

\begin{table*}
\caption{The combinations of input parameters for the models of M-type AGB stars, sorted according to luminosity. The columns list stellar mass $M_\star$, stellar luminosity $L_\star$, effective temperature $T_\star$ of the star, pulsation period $P$, piston velocity amplitude $\Delta u_{\mathrm{p}}$ and the assumed seed particle abundance $n_{\mathrm{gr}}/n_{\mathrm{H}}$. The pulsation period is derived from the period-luminosity relation presented in \cite{fei89}.}             % title of Table
\label{t_grid}      % is used to refer this table in the text
\centering                          % used for centering table
\begin{tabular}{c c r c c c l}        % centered columns
\hline\hline                 % inserts double horizontal lines
Model series  & $M_\star$ & $L_\star$ & $T_\star$ & $P(L_\star)$ & $\Delta u_{p}$ & $\log n_{\mathrm{gr}}/n_{\mathrm{H}}$\\  
 & $[M_{\odot}$] & [$L_{\odot}$]  & [K] &  [d] & [km/s] & [a, b, c, d]\\    
\hline 
L50 & 1 & 5000  & 2600--3000 & 310 & 3.0, 4.0 & $-16.0$, $-15.5$, $-15.0$, $-14.5$\\
L70 & 1 & 7000  & 2600--3100& 395 & 3.0, 4.0 & $-16.0$, $-15.5$, $-15.0$, $-14.5$\\
L10 & 1 & 10000  & 2600--3200 & 525 & 2.0, 3.0 & $-16.0$, $-15.5$, $-15.0$, $-14.5$\\
 \hline
\end{tabular}
\end{table*}

 The variable structures of the atmospheres and winds are produced by simultaneously solving the equations of hydrodynamics, frequency-dependent radiative transfer and time-dependent grain growth, assuming spherical symmetry. The stellar pulsations are simulated by adopting sinusoidal variations of velocity and luminosity at the inner boundary of the atmosphere. When calculating the radiative acceleration of Mg$_2$SiO$_4$ particles, grain-size dependent opacities are used, taking into account both scattering and absorption effects. 

Dust formation is considered as a two step process, starting with the rapid formation of tiny seed nuclei with sizes in the nanometre range, followed the much slower growth of the grains through condensation of material from the gas phase, on time scales comparable to atmosphere and wind dynamics. Currently there is no well-established nucleation theory for oxygen-rich atmospheres, describing the formation of the seed particles. Fundamental questions, e.g. their composition (which may differ from the bulk of material building up the grains) are still a matter of debate. Our modelling of dust formation follows \cite{gail99}, assuming that the growth rate of Mg$_2$SiO$_4$ is determined by the addition of SiO molecules to the grain surface, and using a parameter $n_{\mathrm{gr}}/n_{\mathrm{H}}$ that sets the number of seed particles per hydrogen atom\footnote{In contrast, nucleation in carbon-rich stars is often described by classical nucleation theory, assuming that the seed particle and microscopic grains consist of the same material, i.e. amorphous carbon. The values used here for the parameter $n_{\mathrm{gr}}/n_{\mathrm{H}}$ are comparable to values found in carbon star models.}. These seed particles will start to grow when the thermodynamic conditions are favourable. The equation describing the time-dependent growth of Mg$_2$SiO$_4$ grains is discussed in \cite{bladh13}. Since we assume that the formation of seed particles precedes grain growth (i.e. no new seed nuclei are formed) grain size will be uniform in a given mass layer. Because of changing conditions in the atmosphere and wind, however, grain sizes will differ from layer to layer and change with time.
 
 Each dynamic model is represented by a sequence of snapshots of the radial structures, covering typically hundreds of pulsation periods. From this long sequence we select a series of snapshots, equidistant in phase, during three consecutive pulsation periods. For these snapshots we produce spectra and photometry in an \textit{a posteriori} radiative transfer calculation. The synthetic spectra are computed with opacities from the COMA code \citep{ari00} and mean synthetic photometry is calculated by phase-averaging over the pulsation cycle.

The current study includes 139 solar-mass models of M-type AGB stars, with effective temperatures ranging from 2600\, to 3200\,K. Three different luminosities have been used: $5000\,$L$_{\odot}$, $7000\,$L$_{\odot}$ and $10000\,$L$_{\odot}$. These parameters were chosen to represent common M-type AGB stars, according to stellar evolution models. The pulsation period controlling the variations at the inner boundary is derived from the period-luminosity relation presented in \cite{fei89} and the piston velocity amplitude $\Delta u_{\mathrm{p}}$ ranges between 2-4 km/s, depending on luminosity. This results in shock amplitudes of about 15-20 km/s in the inner atmosphere. Table~\ref{t_grid} shows the combination of input parameters of the models, sorted by luminosity. For each fixed luminosity, the effective temperature, piston velocity amplitude and seed particle abundance are varied according to the values listed.

\section{Wind properties}
\label{s_dyn}

\begin{figure}
\centering
\includegraphics[width=\linewidth]{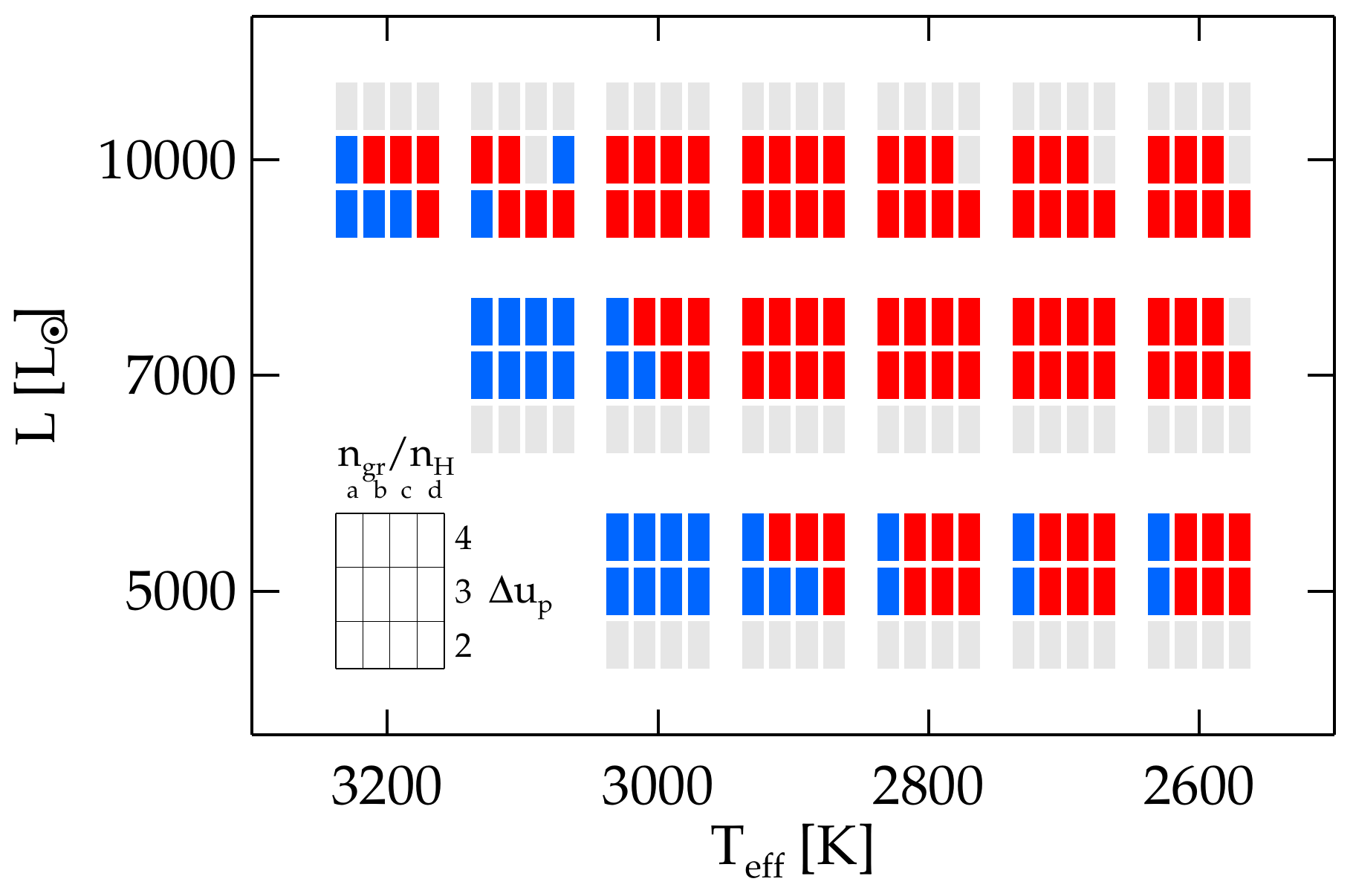}
   \caption{Dynamic behaviour of the models as a function of input parameters. The red rectangles indicate models with a stellar wind, the blue rectangles indicate models where no wind forms, and the grey rectangles indicate combinations of parameters not tested or where the models fail for numerical reason. For each combination of luminosity and effective temperature the seed particle abundance and piston velocity are varied as indicated by the inset box. See Table~\ref{t_grid} for the different values of seed particle abundance $n_{\mathrm{gr}}/n_{\mathrm{H}}$ (a-d).}
    \label{f_ifwind}
\end{figure}

A schematic overview of the dynamical properties of the available models is shown in Fig.\ref{f_ifwind}. The range in effective temperature and luminosity is listed on the x-axis and y-axis, respectively. Each subset consists of twelve boxes and is organised such that the piston velocity is increasing upwards and the seed particle abundance is increasing towards the right. The red boxes represent models that develop a stellar wind, the blue boxes represent models where no wind forms and the grey boxes indicate combinations of parameters not tested or where the models fail for numerical reasons. It is clear from this plot that the dynamical models can produce outflows for a wide range of stellar parameters, although it is generally more difficult to drive winds for high effective temperatures and low luminosities. Judging from these trends, models with lower effective temperature and/or higher luminosity than in this current sample can be expected to produce outflows.\footnote{The current set of models for M-type AGB stars does not lead to any wind models with episodic outflows, whereas such models are found in the grid for C-type AGB stars of \cite{matt10}, as discussed by \cite{erik14}. A possible reason for the different wind behaviour is that dust particles with a large absorption cross-section in the visual and near-IR (e.g. amorphous carbon) cause back-warming effects in the atmosphere. This back-warming can influence the dust formation in the layers below and consequently the dynamics. If the wind-driving dust species is very transparent in the visual and near-IR (e.g. magnesium silicates) and acting predominately through photon scattering, then this feedback through back-warming will not occur.}

\begin{figure}
\centering
\includegraphics[width=\linewidth]{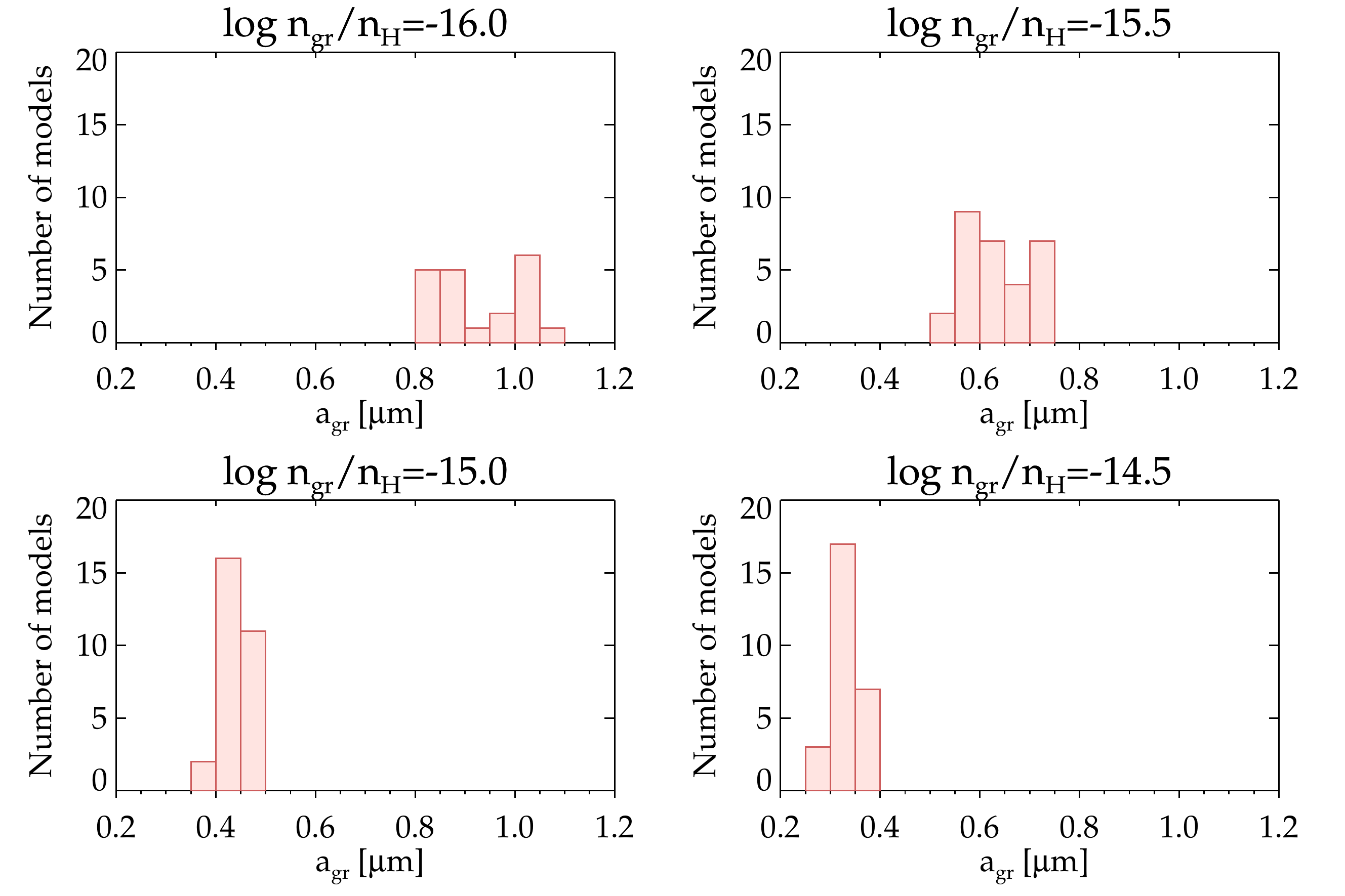}
   \caption{Grain sizes (averaged over pulsation cycle) at the outer boundary of the models that produce a stellar wind, sorted according to different seed particle abundance $n_{\mathrm{gr}}/n_{\mathrm{H}}$.}
    \label{f_gsize}
\qquad\\    
\centering
\includegraphics[width=\linewidth]{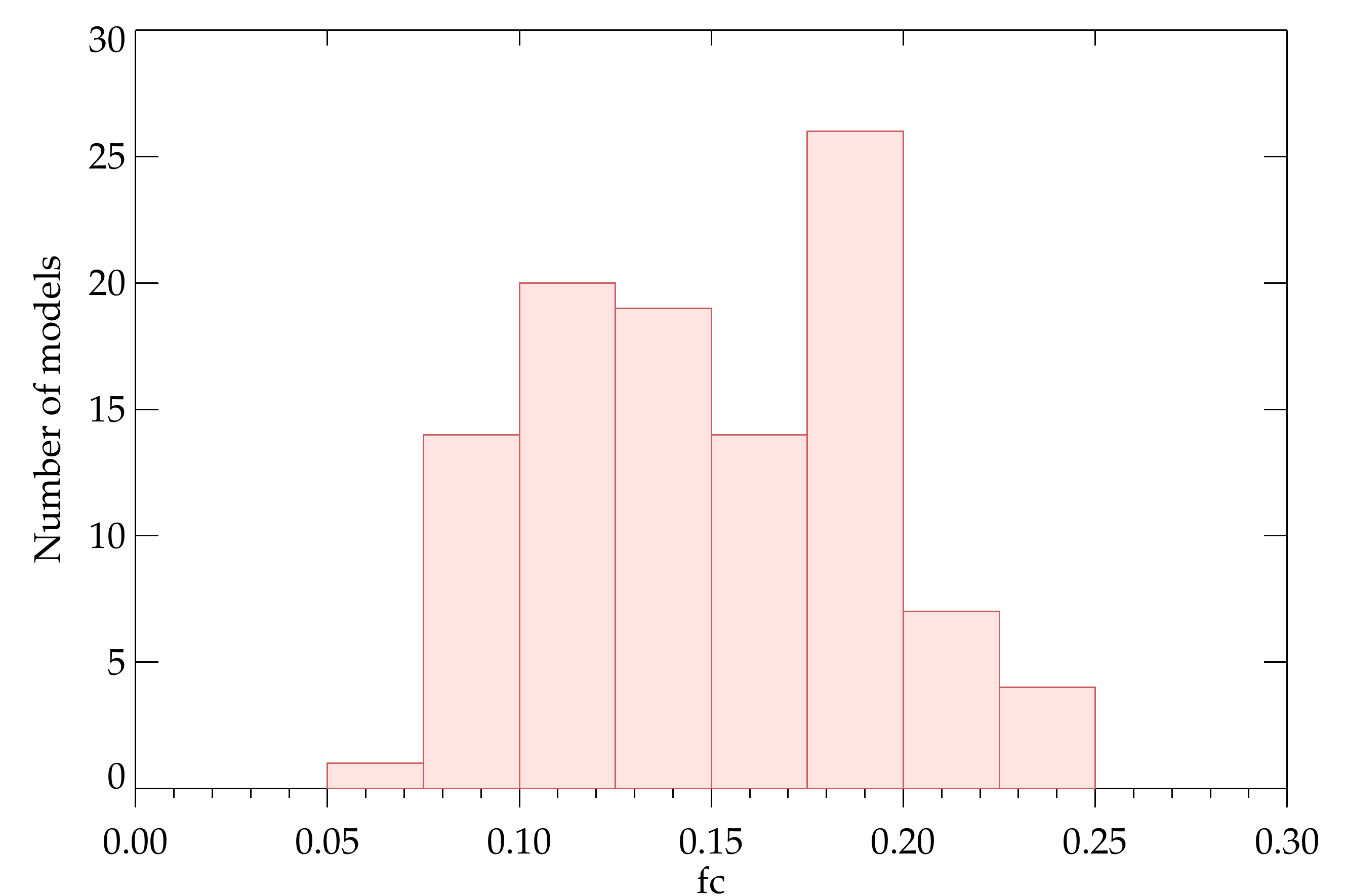}
   \caption{The fraction of Si atoms condensed into grains, averaged over pulsation cycle at the outer boundary of the models that produce a stellar wind.}
    \label{f_fc}
\end{figure}

While the effects of certain parameters are rather straight-forward ($L_*$, $T_*$), or have been discussed in other papers ($\Delta u_{\mathrm{p}}$), the seed particle abundance needs some further discussion. The range of values for this parameter is limited on both ends since photon scattering as a wind-driving mechanism requires grains in a certain size range \citep[see][for a discussion]{hof08toy}. Fewer seeds make the growth of individual grains more efficient (less competition for material), but may actually lead to a smaller collective opacity if the grains grow beyond the optimal size range. More seeds lead to stronger competition for condensable material, resulting in smaller grains. This effect is demonstrated in Fig.~\ref{f_gsize}, showing the grain radius averaged over pulsation cycle in the outermost layers of the dynamical models that produce a wind, sorted according to seed particle abundance. The degrees of condensed silicon (Si) in the models that produce outflows are shown in Fig.~\ref{f_fc}, with typical values ranging between 10-25\%, corresponding to dust-to-gas mass ratios of about $10^{-3}$. This is in reasonable agreement with recent results by \cite{khou14} for W Hya, who find that about a one third of Si is condensed into grains. According to our models the degree of condensed magnesium will be about twice as much as for Si since Mg$_2$SiO$_4$ has two magnesium atoms per silicon atom and the element abundances of Si and Mg are comparable.

\begin{figure*}
\centering
\includegraphics[width=9cm]{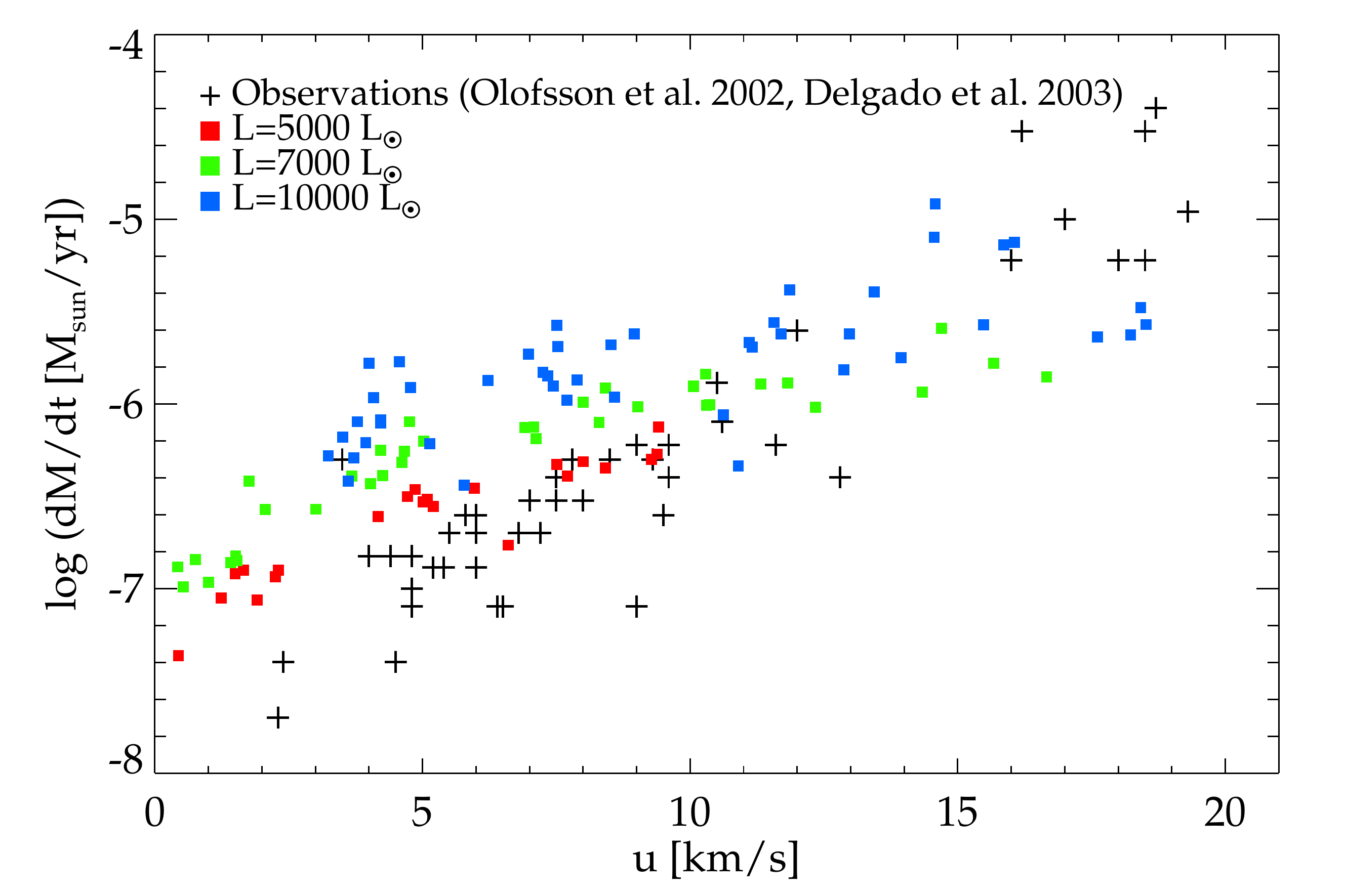} 
\includegraphics[width=9cm]{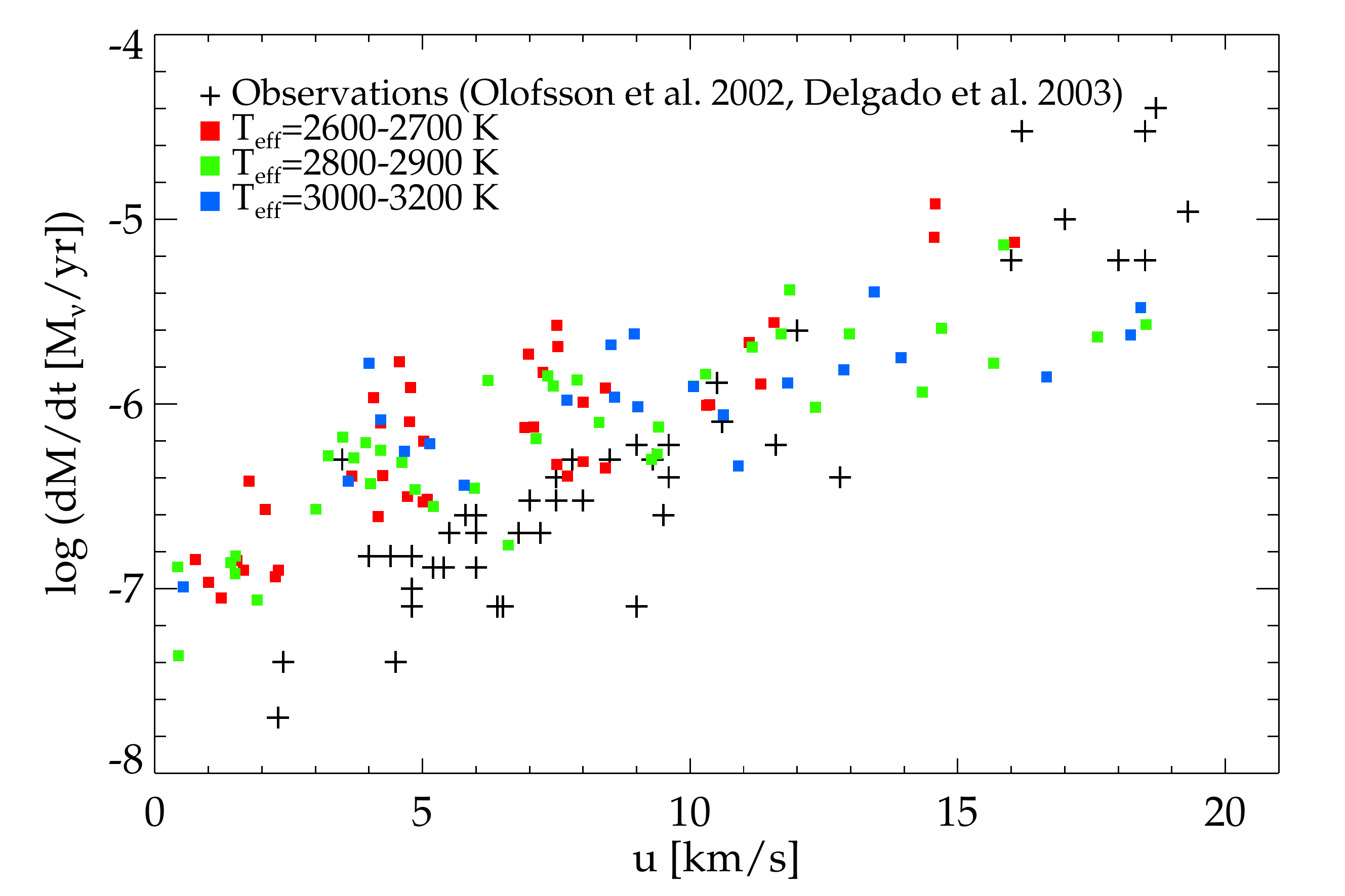}
\includegraphics[width=9cm]{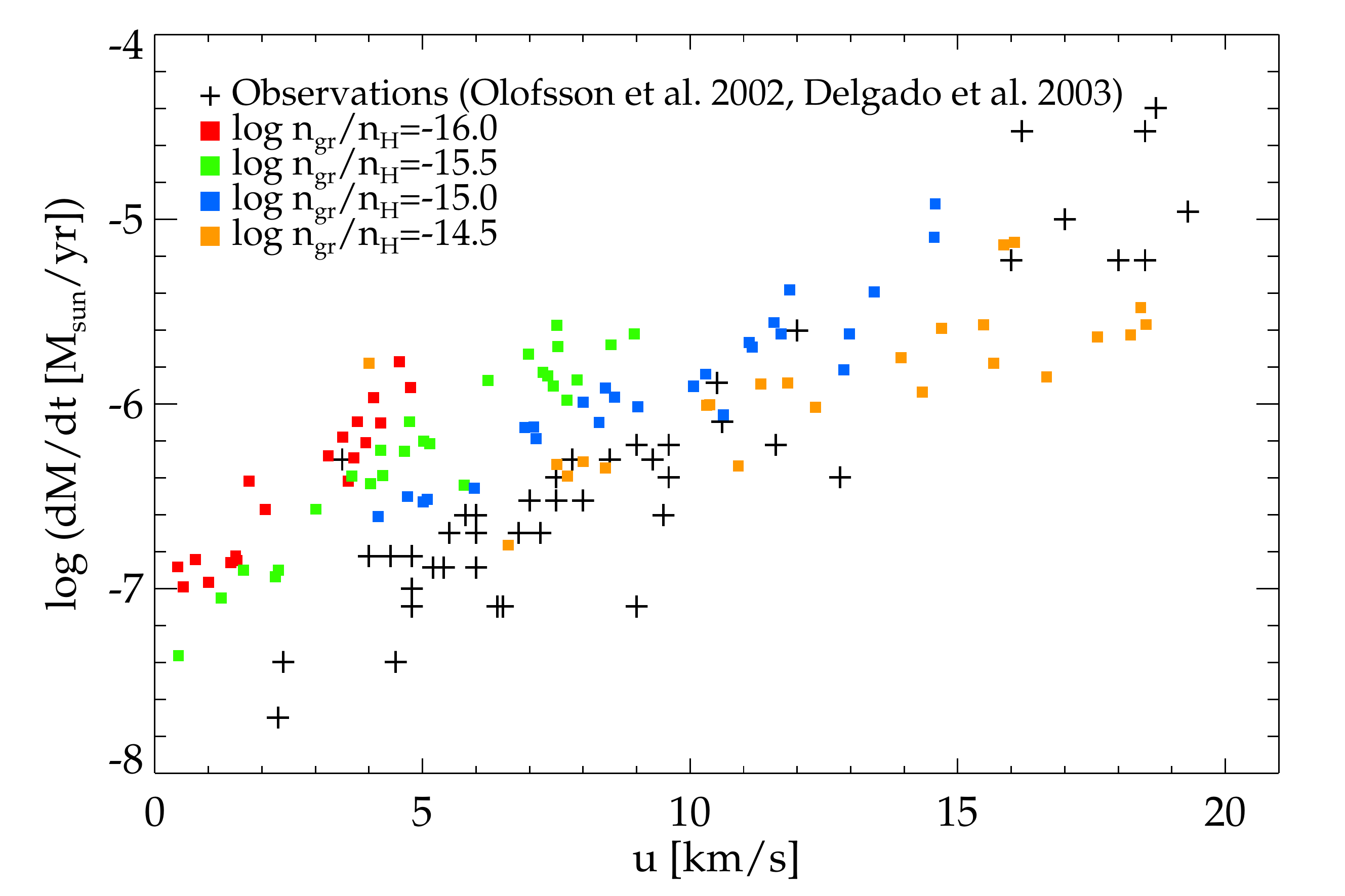} 
\includegraphics[width=9cm]{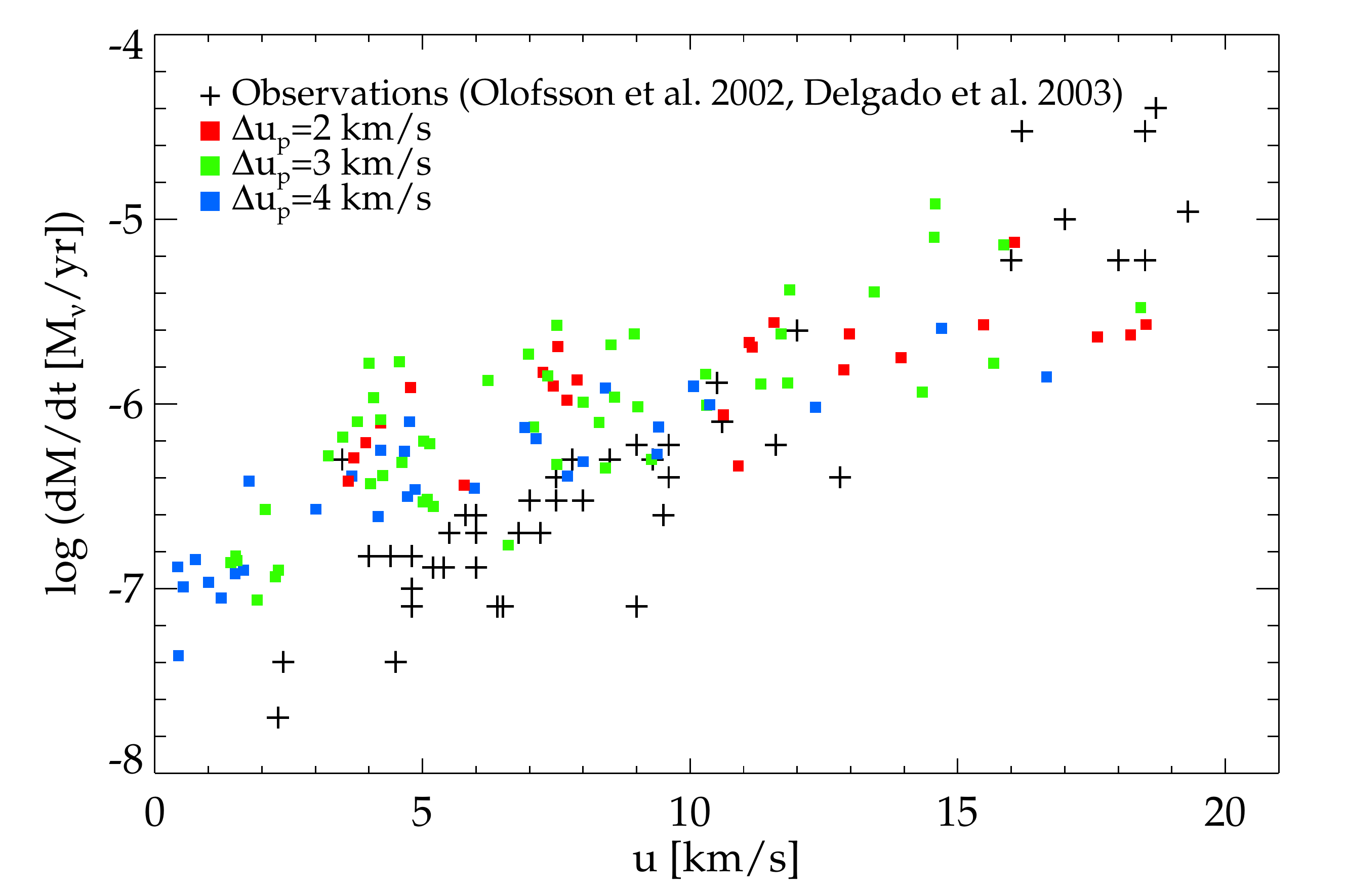}
   \caption{Observed mass-loss rates vs. wind velocities of M-type AGB stars \citep[][plus signs]{hans02,gondel03} and the corresponding properties of all models which develop a wind (indicated in red in Fig.~\ref{f_ifwind}). All panels show the same data but with different colour coding according to stellar luminosity (upper left), effective temperature (upper right), seed particle abundance (lower left) and piston velocity amplitude (lower right). The stellar mass is one solar mass for all models.}
    \label{f_dyn2}
\end{figure*}

A comparison of wind velocities and mass-loss rates of the models with corresponding empirical data for M-type AGB stars, derived from observations of several CO-lines, is shown in Fig.~\ref{f_dyn2}. The agreement with observations is good, not the least considering that this set of models does not correspond to a stellar population, but rather a grid of input parameters where not all combinations of parameters are equally likely. Both high and low mass-loss rates are reproduced, and higher mass-loss rates could probably be reached by increasing the stellar luminosity of the wind models further. 

The four panels of Fig.~\ref{f_dyn2} show the same data but the models are colour-coded in different ways to illustrate the effects of different input parameters. The top left panel of Fig. \ref{f_dyn2} shows the dynamical properties of the wind models, colour-coded according to luminosity. The models with different values for the luminosity form bands where higher luminosity correlates with higher mass-loss rates. If we instead plot the dynamical properties colour-coded by seed particle abundance (lower left panel), the models form bands correlating more with wind velocity. In right panels of Fig. \ref{f_dyn2} we also plot the models sorted according to effective temperatures and piston velocities, but we see no distinct trends in wind properties with respect to these parameters.

\section{Visual and near-IR photometry}
\label{s_nir}

Dust-driven winds can be accelerated by both absorption and scattering of stellar photons on dust grains, but scattering is only significant if the grains grow to adequate sizes. If true absorption is the dominant source of the radiative acceleration then significant circumstellar reddening will occur; the dusty envelope will veil the molecular features in the visual wavelength region and emit a significant infrared excess.  If scattering is the dominant source of the momentum there will be much less circumstellar reddening; changes in the molecular features during the pulsation cycle will be visible at visual wavelengths and there will not be much infrared excess, even though dust is present. Since the visual and near-IR spectra are strongly sensitive to the wind-driving mechanism, i.e. the optical properties of the wind-driving dust species, we use the colours $(J-K)$ and $(V-K)$ when comparing model results with observational data. The sources of the observational data sets used here are described in detail in \cite{bladh13}. 

\begin{table}
\caption{Input parameters for the dynamical models resulting in the photometric variations shown in Fig.~\ref{f_grid4}.}             % title of Table
\label{t_grid2}      % is used to refer this table in the text
\centering                          % used for centering table
\begin{tabular}{c c c c c}        % centered columns
\hline\hline                 % inserts double horizontal lines
$L_\star$ & $T_\star$ & $\Delta u_{p}$ & $\log n_{\mathrm{gr}}/n_{\mathrm{H}}$ & Line\\  
 $L_{\odot}$  & [K] & [km/s] &  & colour \\    
\hline 
5000  & 2600 & 4.0 & $-15.0$ & red\\
5000  & 2800 & 4.0 & $-15.0$ & red\\\
7000  & 2600 & 3.0 & $-15.0$ & green\\
7000  & 2800 & 3.0 & $-15.0$ & green\\
7000  & 3000 & 3.0 & $-15.0$ & green\\
10000  & 2600 & 2.0 & $-15.0$ & blue\\
10000  & 2800 & 2.0 & $-15.0$ & blue\\
10000  & 3000 & 2.0 & $-15.0$ & blue\\
10000  & 3000 & 3.0 & $-15.5$ & blue\\
10000  & 3200 & 2.0 & $-15.5$ & blue\\
 \hline
\end{tabular}
\end{table}

\begin{figure}
\centering
\includegraphics[width=\linewidth]{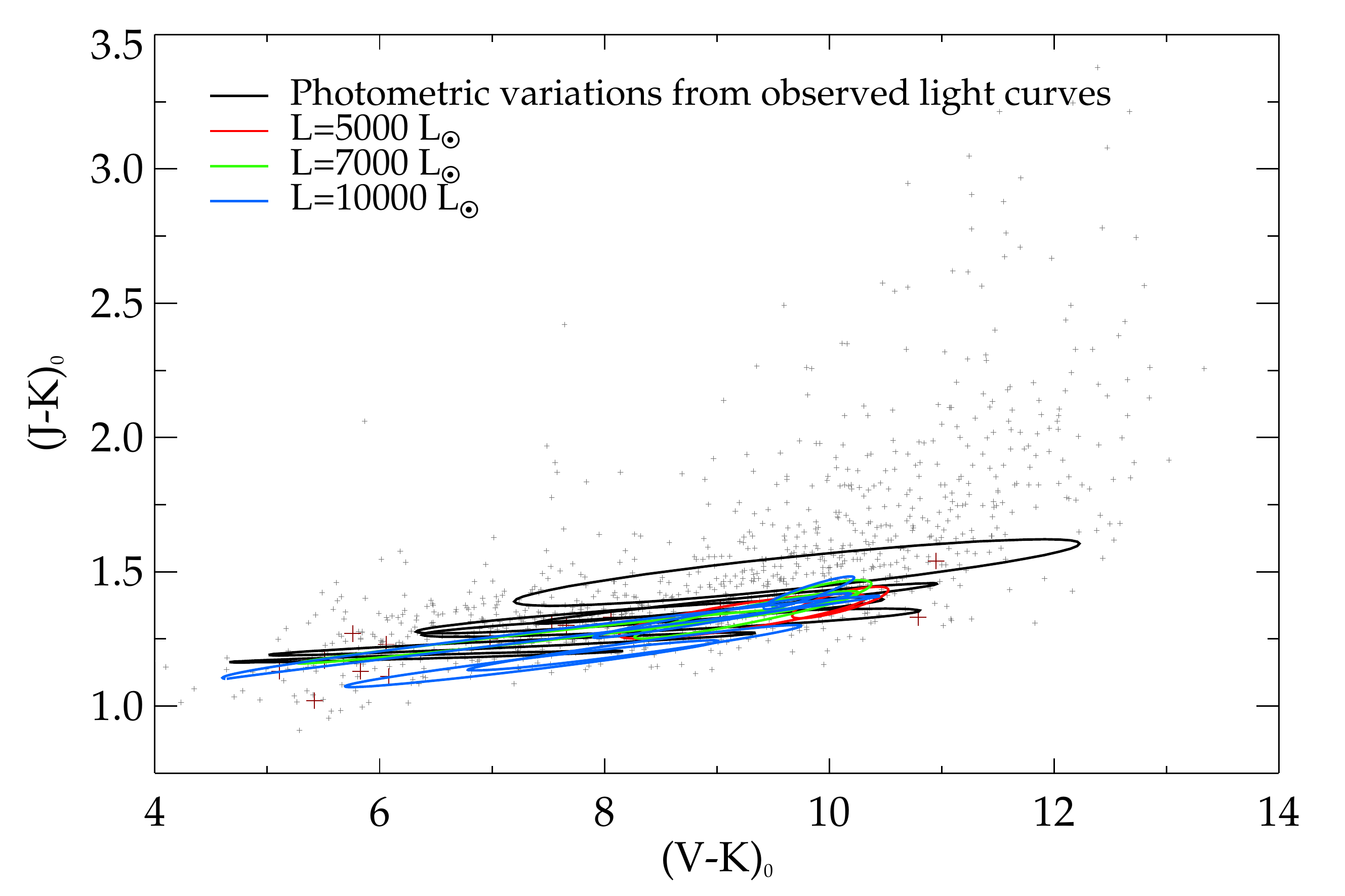} 
\includegraphics[width=\linewidth]{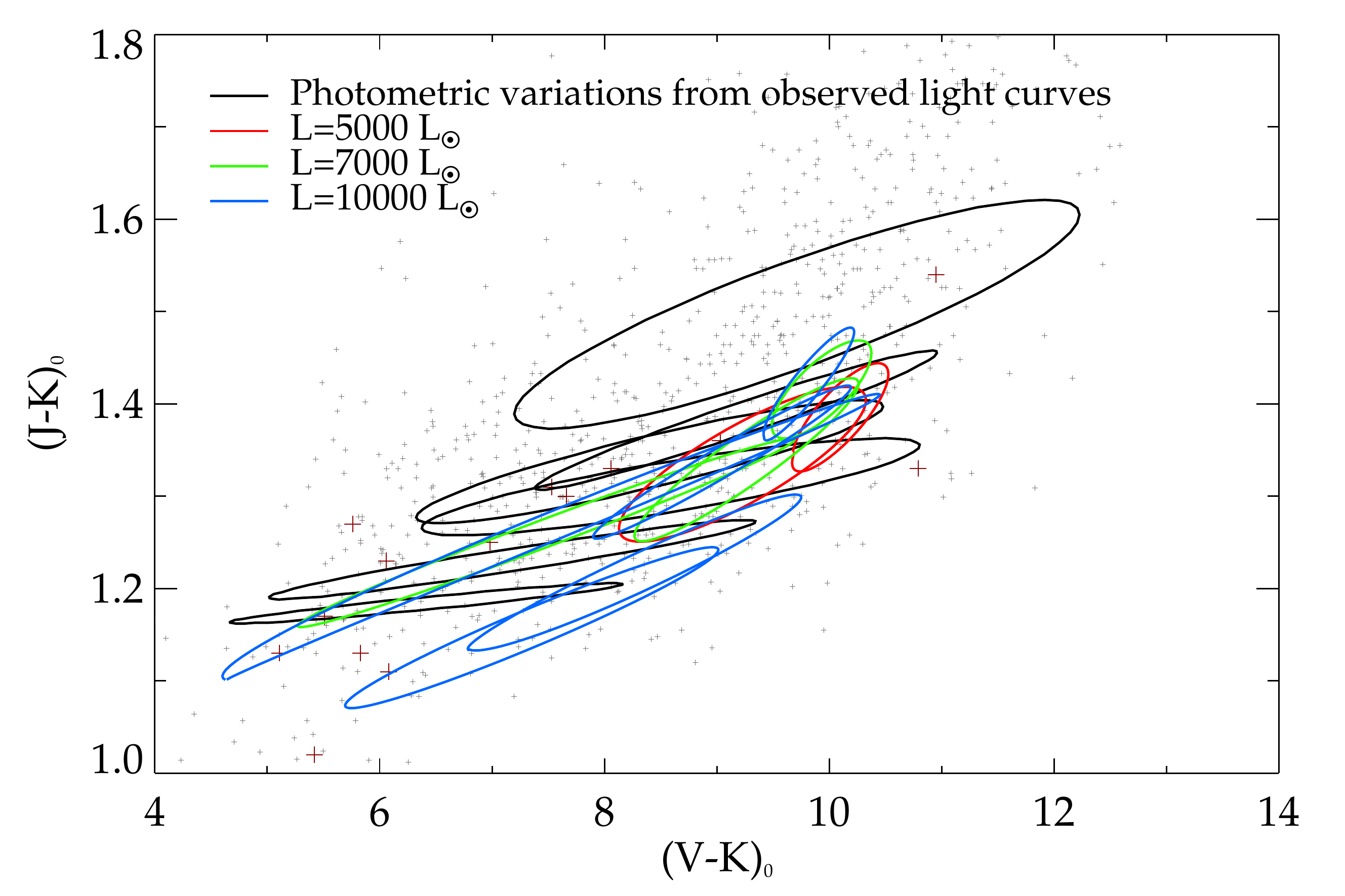} 
   \caption{Photometric variations during a pulsation cycle in $(J-K)$ vs. $(V-K)$. The black loops show photometric variations derived from sine fits of observed light-curves for a set of M-type miras (R Car, R Hya, R Oct, R Vir, RR Sco, T Col and T Hor) and the coloured loops show synthetic photometric variation from a few selected models (see Table~\ref{t_grid2}). For the observed miras we adopted photometric data in the visual from \citet{Eggen75a} and \citet{men67} and complemented those with the near-IR data published by \citet{WhiMF00}. The synthetic loops are calculated from sine fits of the light-curves in the same way as the observational data. See \cite{bladh13} for more details concerning the sinefit of observed light curves. The bottom panel shows the same data as the top panel, but zoomed in. The observations in grey (plus signs) are presented in Fig.~\ref{f_grid2}.}
    \label{f_grid4}
\centering
\quad\\
\includegraphics[width=\linewidth]{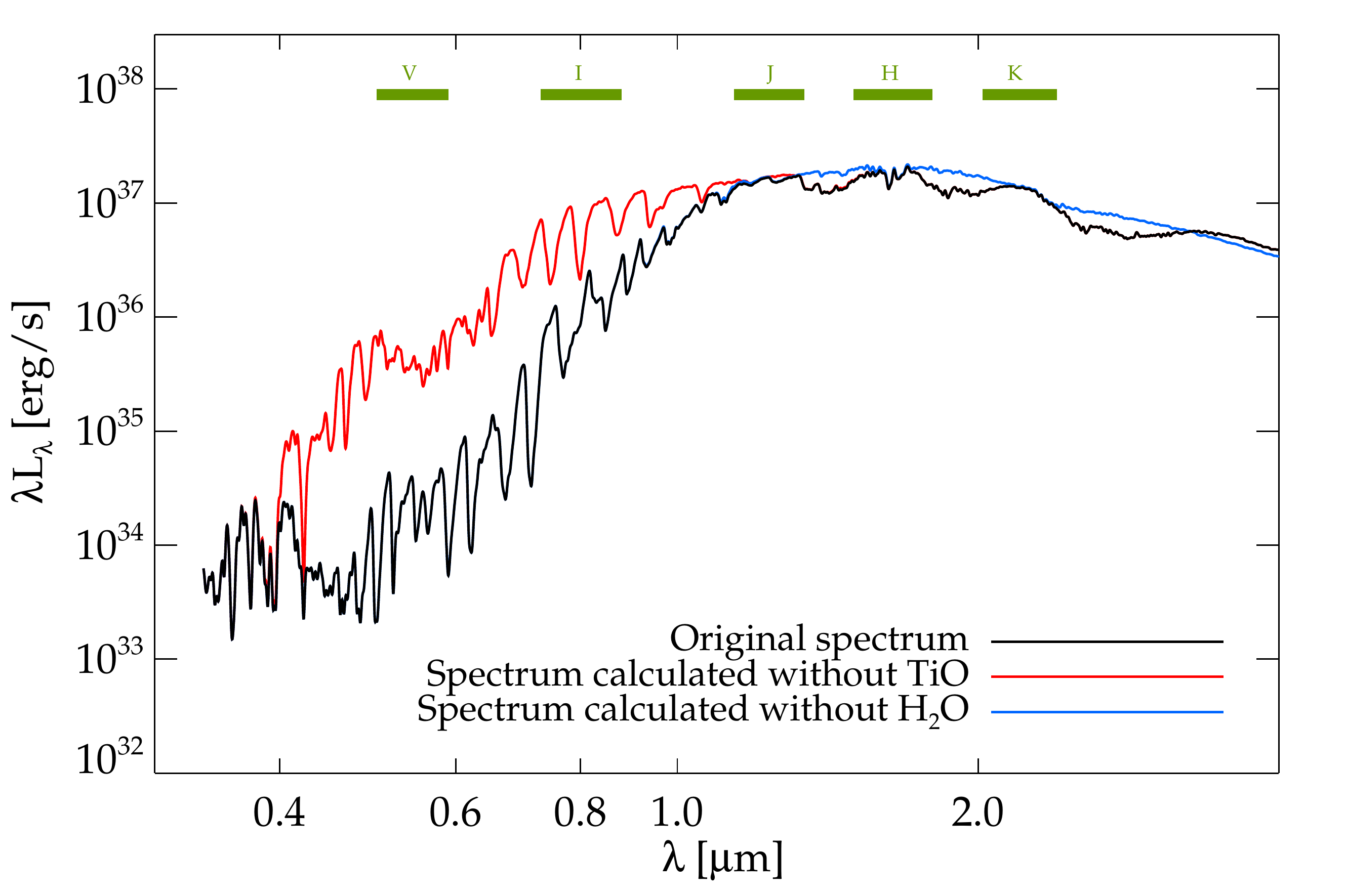} 
   \caption{Spectral energy distribution as a function of wavelength for a model with input parameters \mbox{$M=1\,\mathrm{M}_{\odot}$}, \mbox{$L=5000\,\mathrm{L}_{\odot}$}, \mbox{$T_{*}=2800\,$K}, \mbox{$u_\mathrm{p}=4\,$km/s}, and \mbox{$\log n_{\mathrm{gr}}/n_{\mathrm{H}}=14.5$}. The original spectrum is shown in black and the coloured curves show spectra calculated without TiO opacities (red) and H$_2$O opacities (blue), indicating which photometric bands will be affected by changes in the corresponding features.}
    \label{f_spectio}
\end{figure}

AGB stars are variable stars with photometric properties that change with phase. Realistic dynamical models should be able to reproduce this photometric variation during the pulsation cycle.  In Fig.~\ref{f_grid4} we plot the synthetic colour $(J-K)$ vs. $(V-K)$ for a few selected models during a pulsation cycle, together with phase-dependent colours constructed from sine fits of observed light curves \citep[see][for more details]{bladh13}. These photometric variations form loops in the colour-colour diagram, with different positions, shapes and tilts. For M-type AGB stars, both the synthetic and observed colour loops are characterised by large variations in $(V-K)$ and small variations in $(J-K)$, as can be seen in Fig.~\ref{f_grid4}. The colour $(V-K)$ reaches its peak value during luminosity minimum, followed by a trough during luminosity maximum. A closer examination of the synthetic spectra reveals that the large variations in $(V-K)$ during a pulsation cycle are caused by a change in the abundance of molecules (TiO, see Fig.~\ref{f_spectio}), and not by changes in the amount of dust present \citep{bladh13}. This is a strong indication that the wind-driving dust species in M-type AGB stars are quite transparent in the visual and near-IR, otherwise these variations would not be dominated by molecular features.

\begin{figure*}
\centering
\includegraphics[width=9cm]{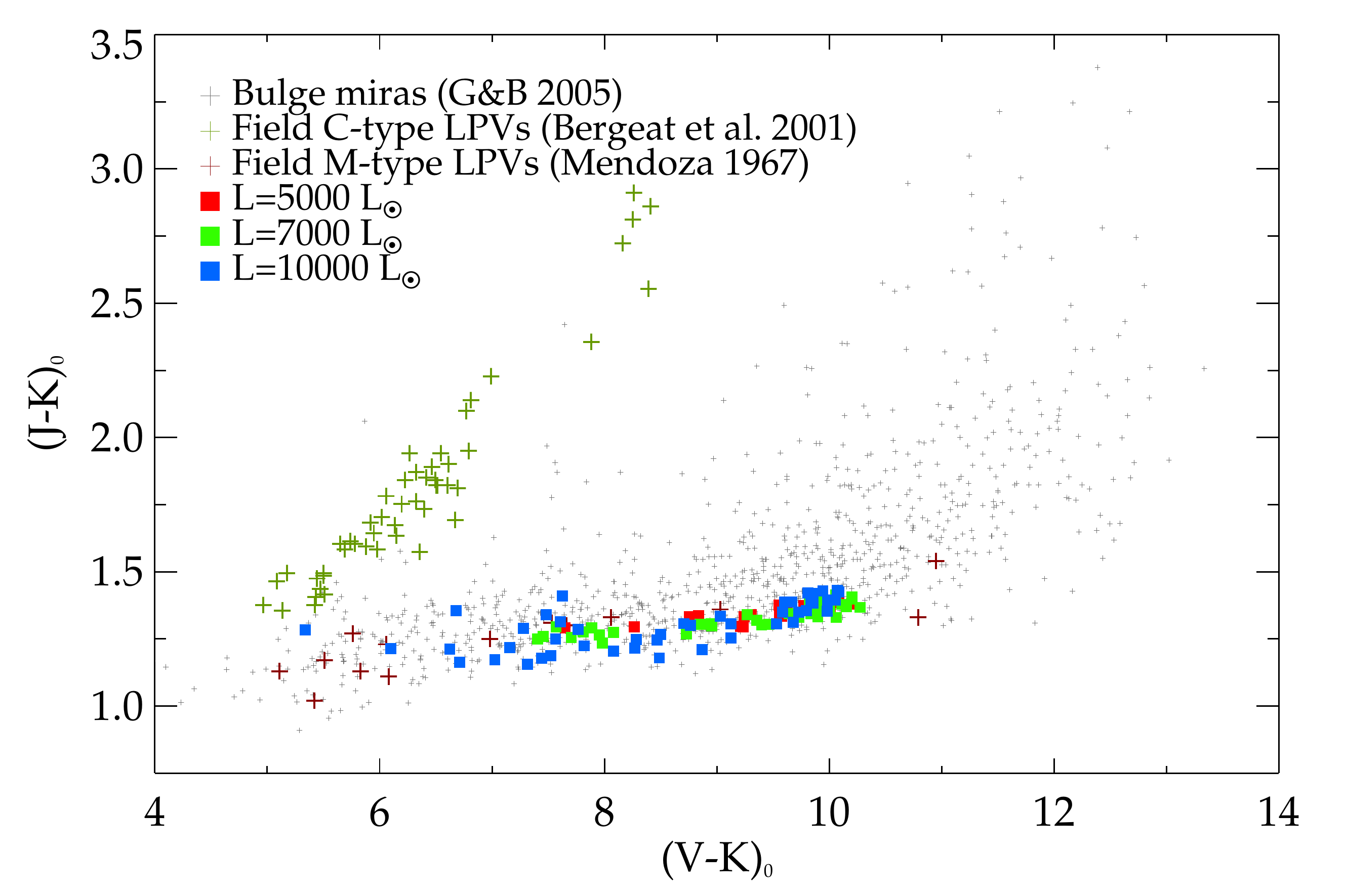} 
\includegraphics[width=9cm]{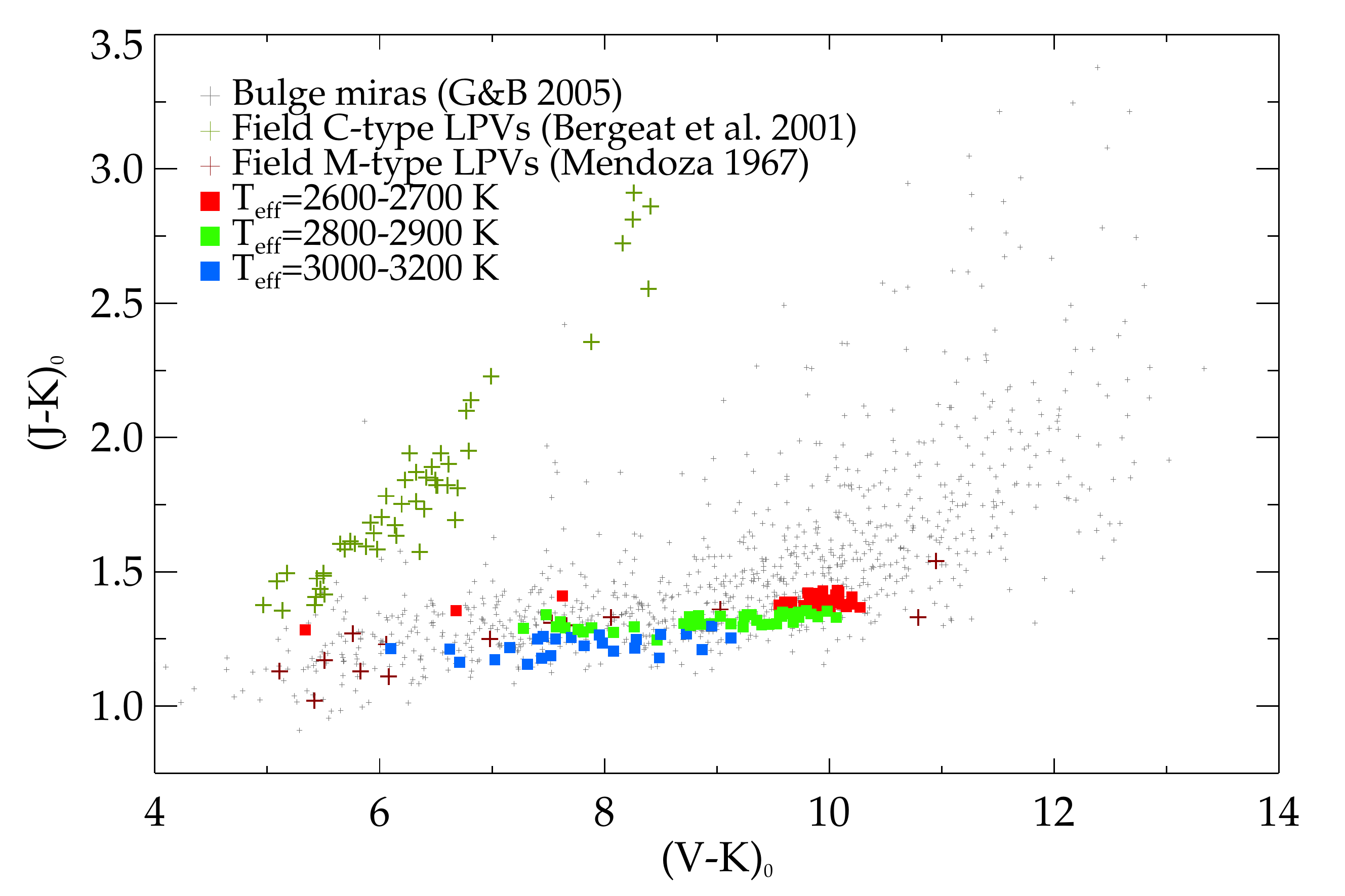} 
\includegraphics[width=9cm]{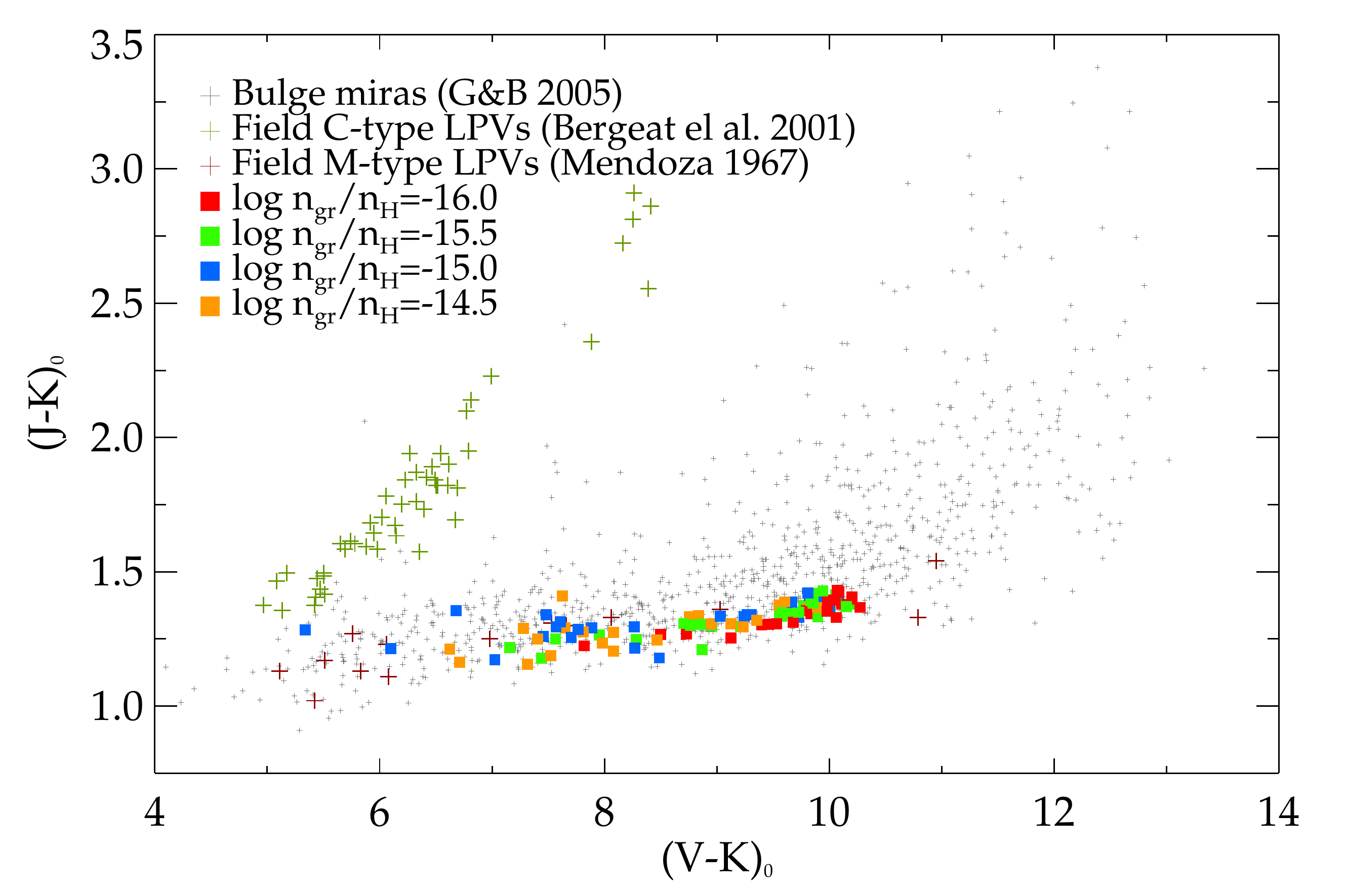} 
\includegraphics[width=9cm]{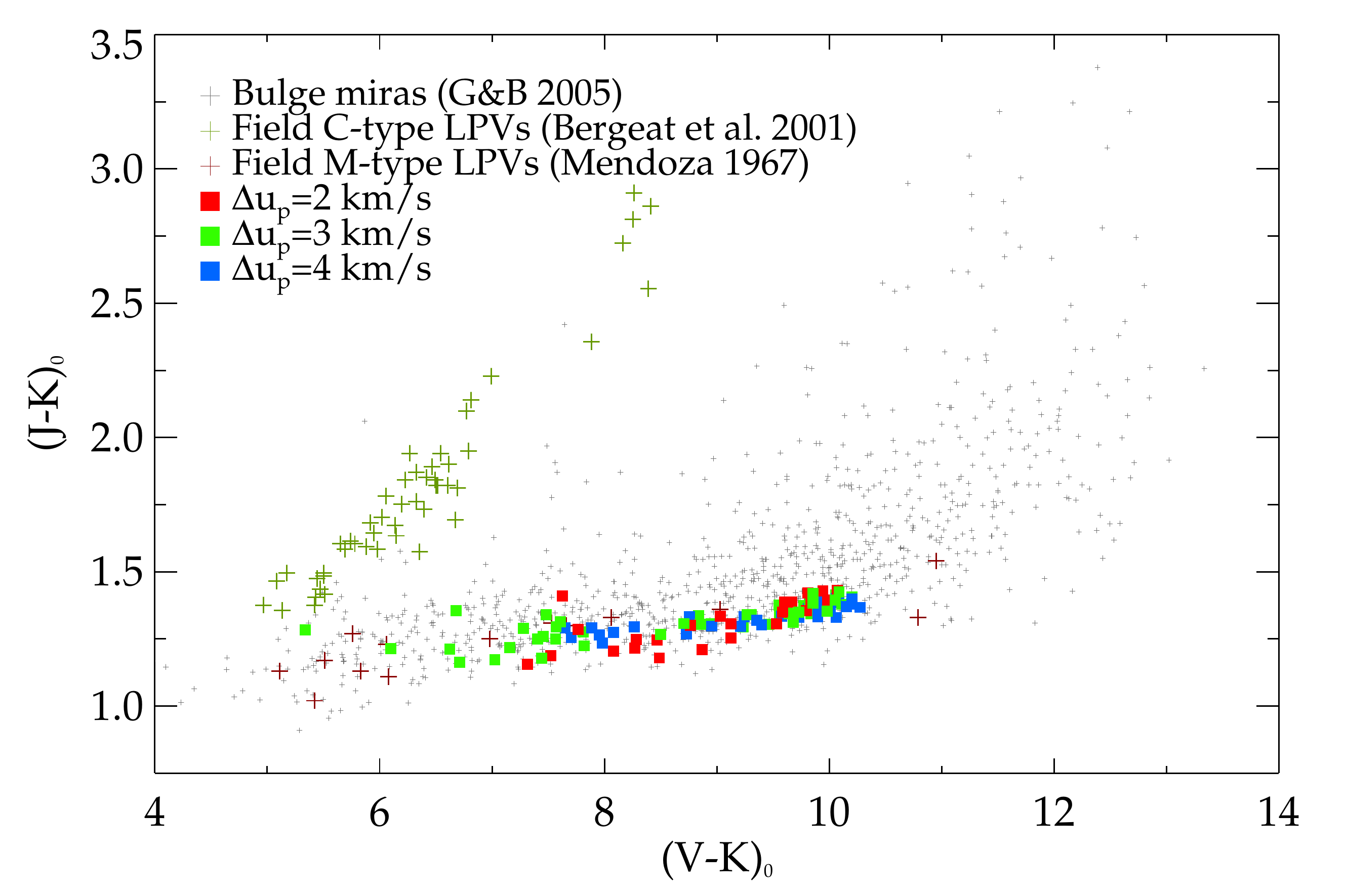} 
   \caption{Observed and mean synthetic colours $(J-K)$ vs. $(V-K)$. The observational data (plus signs) is compiled from different sources: Galactic Bulge miras \citep[grey]{gbmiras}, field M-type LPVs \citep[red]{men67}, and C-rich giants \citep[green]{b01cstars}. The model data (filled squares) is from models which develop a wind, as indicated in red in Fig.~\ref{f_ifwind}. All panels show the same data but with different colour coding according to stellar luminosity (upper left), effective temperature (upper right), seed particle abundance (lower left) and piston velocity (lower right). The stellar mass is one solar mass for all models. Most of the observational data are single epoch measurements, whereas the synthetic colours are means over the pulsation cycle.}
    \label{f_grid2}
\end{figure*}

Looking at Fig.~\ref{f_grid4}, it is clear that phase-averaged colours will be affected by both the position of the loops in the colour-colour diagram and the variations in colour during a pulsation cycle. Fig.~\ref{f_grid2} shows phase-averaged synthetic colours (filled squares), as well as observed colours for both M-type (grey and red plus signs) and C-type AGB stars (green plus signs). The synthetic colours are characterised by blue values in $(J-K)$, especially compared to values of C-type AGB stars, with small differences between the individual dynamical models. There is a strong correlation between the effective temperatures of the models and the values of $(J-K)$, as can be seen in the top right panel of Fig.~\ref{f_grid2}. This is probably because the temperature of the star has a strong impact on the water features \mbox{(see Fig.~\ref{f_spectio})}.

The phase-averaged synthetic $(V-K)$ colours show large differences between individual models. The values of the mean $(V-K)$ colours are mostly determined by the range of variation during the pulsation cycle, i.e. how far the loops extend to the blue side of the diagram. As mentioned above, the variation in visual fluxes is determined by changes in TiO, which in turn reflects changes in temperature during the pulsation cycle. The stronger the temperature variations, the wider the loops, resulting in bluer average $(V-K)$ colours.

The conclusion we can draw from Fig.~\ref{f_grid2} is that the mean synthetic colours reproduce well the observed colours from field M-type AGB stars \citep{men67} and the bulk of the observed values from Bulge miras \citep{gbmiras}, especially the large spread in $(V-K)$. The observed colours from Bulge miras show a much larger spread in $(J-K)$ than the synthetic colours, but this could partly be due to metallicity effects and the fact that we assume solar abundances of C and O in the models. A higher C/O-ratio (closer to 1 than the 0.48 that we assume) should give redder values of $(V-K)$ and $(J-K)$, judging from hydrostatic model atmospheres.

\section{Mid-IR spectra}
\label{s_mir}

An issue with the current wind models is that, although they produce dynamical properties and photometry in the visual and near-IR in agreement with observations (see Secs.~\ref{s_dyn} and \ref{s_nir}), they do not show the prominent silicate features at 10~$\mu$m and 18~$\mu$m that are observed in many oxygen-rich AGB stars. These silicate features have been identified as belonging to magnesium-iron silicates, such as olivine ([Mg,Fe]$_2$SiO$_4$) and pyroxene ([Mg,Fe]SiO$_3$), and are caused by stretching and bending resonances in the SiO$_4$-tetrahedron.

\begin{figure}
\centering
\includegraphics[width=\linewidth]{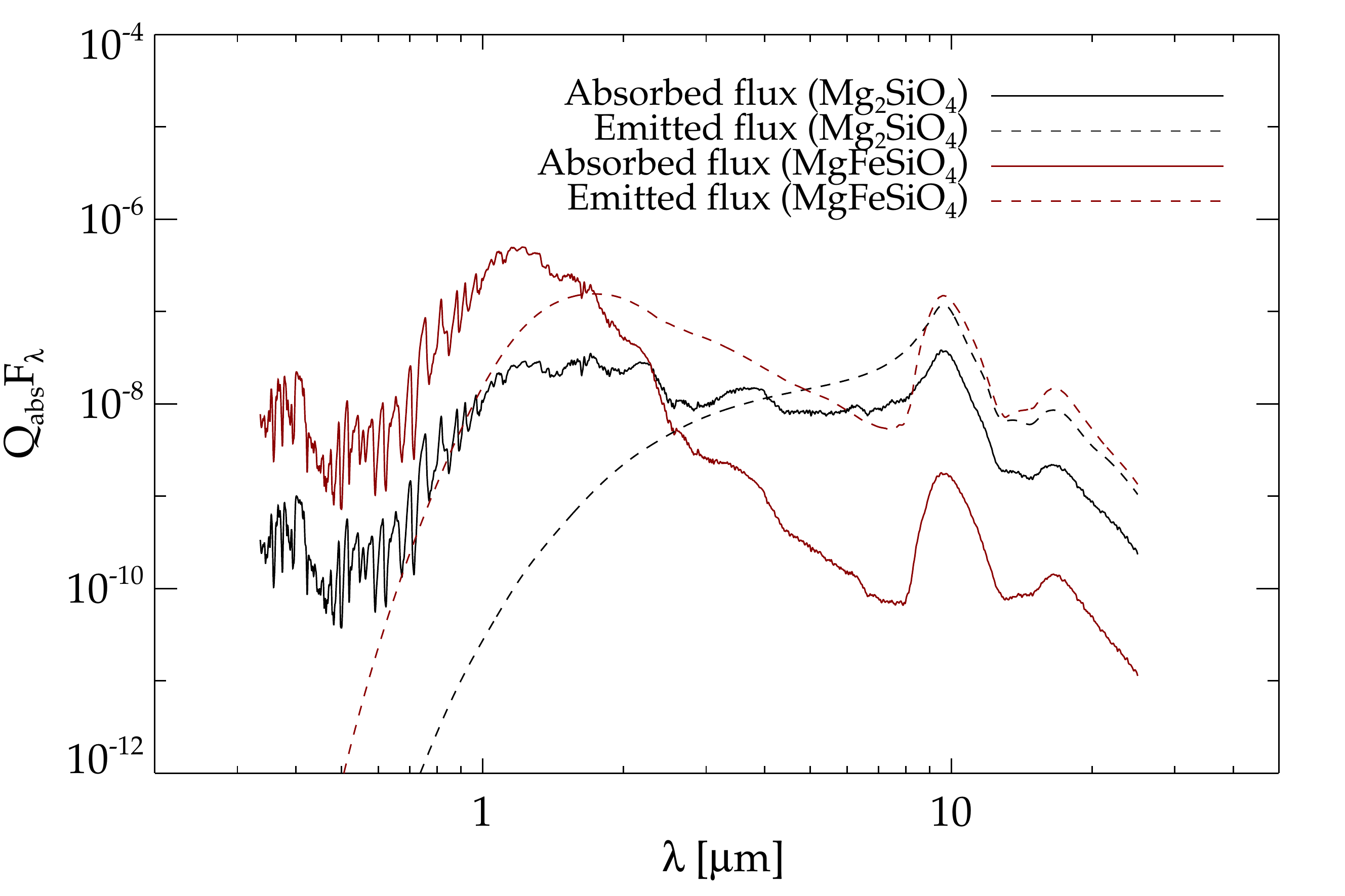}
   \caption{Absorbed and emitted flux, $Q_{\mathrm{abs}}(\lambda)F_{\lambda}$ and $Q_{\mathrm{abs}}(\lambda)B_{\lambda}$, for two different grain materials, Mg$_2$SiO$_4$ (black) and MgFeSiO$_4$ (red). The absorbed light is calculated using a geometrically diluted radiation field from a hydrostatic atmosphere (at 2\,R$_*$ for Mg$_2$SiO$_4$ and at 10\,R$_*$ for MgFeSiO$_4$) with stellar parameters typical of an M-type AGB star \mbox{($M=1\,\mathrm{M}_{\odot}$, $L=5000\,\mathrm{L}_{\odot}$ and $T_{*}=2800\,$K)}. The emitted light is calculated using a Planckian source function at a grain temperature of \mbox{$T_{\mathrm{d}}=1000\,$K}. The optical data for Mg$_2$SiO$_4$ is from \cite{jag03} and the optical data for MgFeSiO$_4$ is from \cite{dor95}. We note that the local radiation field in the wind-acceleration zone will be more pronounced in the mid-IR region than what is shown here because of the addition of the emitted light from the dust component.}
    \label{f_flux}
\end{figure}

The optical properties of Fe-free silicates offer an explanation as to why the synthetic spectra do not show these prominent features. In the wavelength region where AGB stars emit most of their flux the absorption cross-section of magnesium silicates is very low, in particular compared to the spectral features at 10~$\mu$m and 18~$\mu$m, so the heating of the grains by stellar light is moderately efficient. \mbox{Fig. \ref{f_flux}} shows the absorbed and emitted light for Mg$_2$SiO$_4$ grains (black curves), assuming a geometrically diluted radiative flux from a hydrostatic atmosphere (with stellar parameters typical of an M-type AGB star \mbox{$M=1\,\mathrm{M}_{\odot}$, $L=5000\,\mathrm{L}_{\odot}$ and $T_{*}=2800\,$K)} and a dust temperature of \mbox{$T_{\mathrm{d}}=1000\,$K}. The energy balance is set mostly in the mid-IR wavelength region and the grains cool efficiently. This is the reason why such grains can form close to the stellar surface, but it also leads to a rapidly falling dust temperature in the wind.
 
In order to produce silicate features, the grain temperature needs to fall less rapidly with distance from the star than for pure Mg$_2$SiO$_4$ grains. Fig. \ref{f_flux} indicates that the inclusion of Fe in the silicates will help to achieve this; for MgFeSiO$_4$ grains (red curves) light is mostly absorbed in the near-IR wavelength region and then emitted in mid-IR region. However, because of the increased near-IR absorption Fe-bearing silicates can only become thermally stable further away from the star. Simple estimates and results from dynamical models with frequency-dependent radiative transfer predict that silicates with significant Fe content (e.g. MgFeSiO$_4$) become stable at \mbox{$5-10\,R_*$} \citep{woi06fe,hof08toy,bladh12}. \cite{woi06fe} investigated the growth of inhomogeneous dust grains, composed of a mixture of Mg$_2$SiO$_4$, SiO$_2$, Al$_2$O$_3$, TiO$_2$, and solid Fe in a stellar outflow. He found that once Fe inclusion is possible, it will control the temperature of the inhomogeneous dust grains and act as a thermostat, keeping the grain temperature just below the sublimation limit. The amount of included Fe is thereby adjusted in a self-regulating process.

\begin{figure}
\centering
\includegraphics[width=\linewidth]{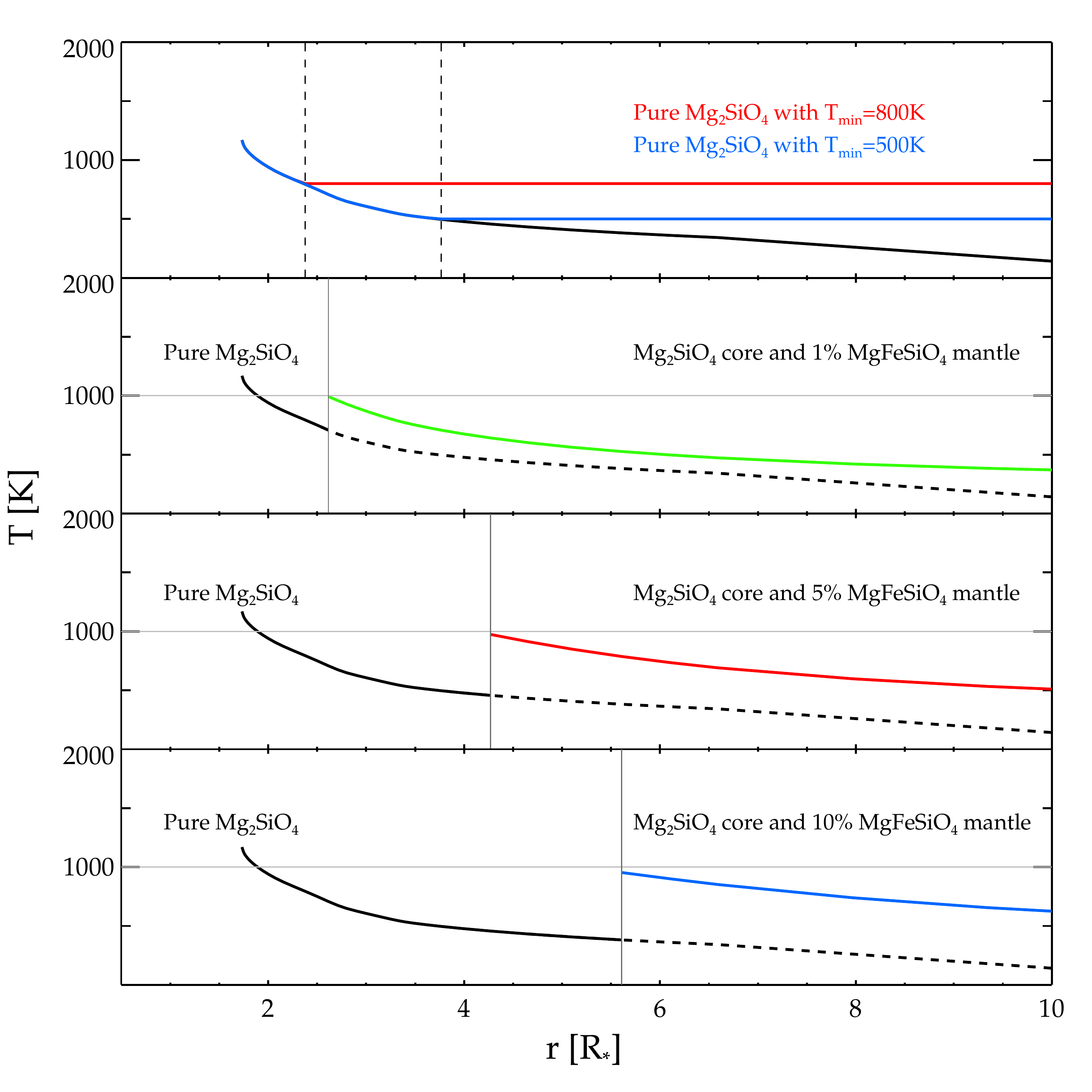}
   \caption{Grain temperature as a function of distance from the star for a model with stellar parameters \mbox{$M=1\,\mathrm{M}_{\odot}$}, \mbox{$L=5000\,\mathrm{L}_{\odot}$} and \mbox{$T_{*}=2800\,$K} during luminosity minimum. \textit{Top panel:} The original temperature structure of the Mg$_2$SiO$_4$ grains (black) and the temperature structure when setting a minimum temperature of 800\,K (red) and 500\,K (blue). \textit{Lower panels:} The original temperature structure of the Mg$_2$SiO$_4$ grains (black) and the temperature structure of dust particles with a core of  Mg$_2$SiO$_4$ and a mantle of MgFeSiO$_4$ when it is thermally stable ($T<1000\,$K), as indicated by the vertical grey line. The green, red and blue lines show the temperature structure for a mantle size that is 1\%, 5\% and 10\% of the grain radius, respectively.}
    \label{f_temp}
\end{figure}

\begin{figure}
\centering
\includegraphics[width=\linewidth]{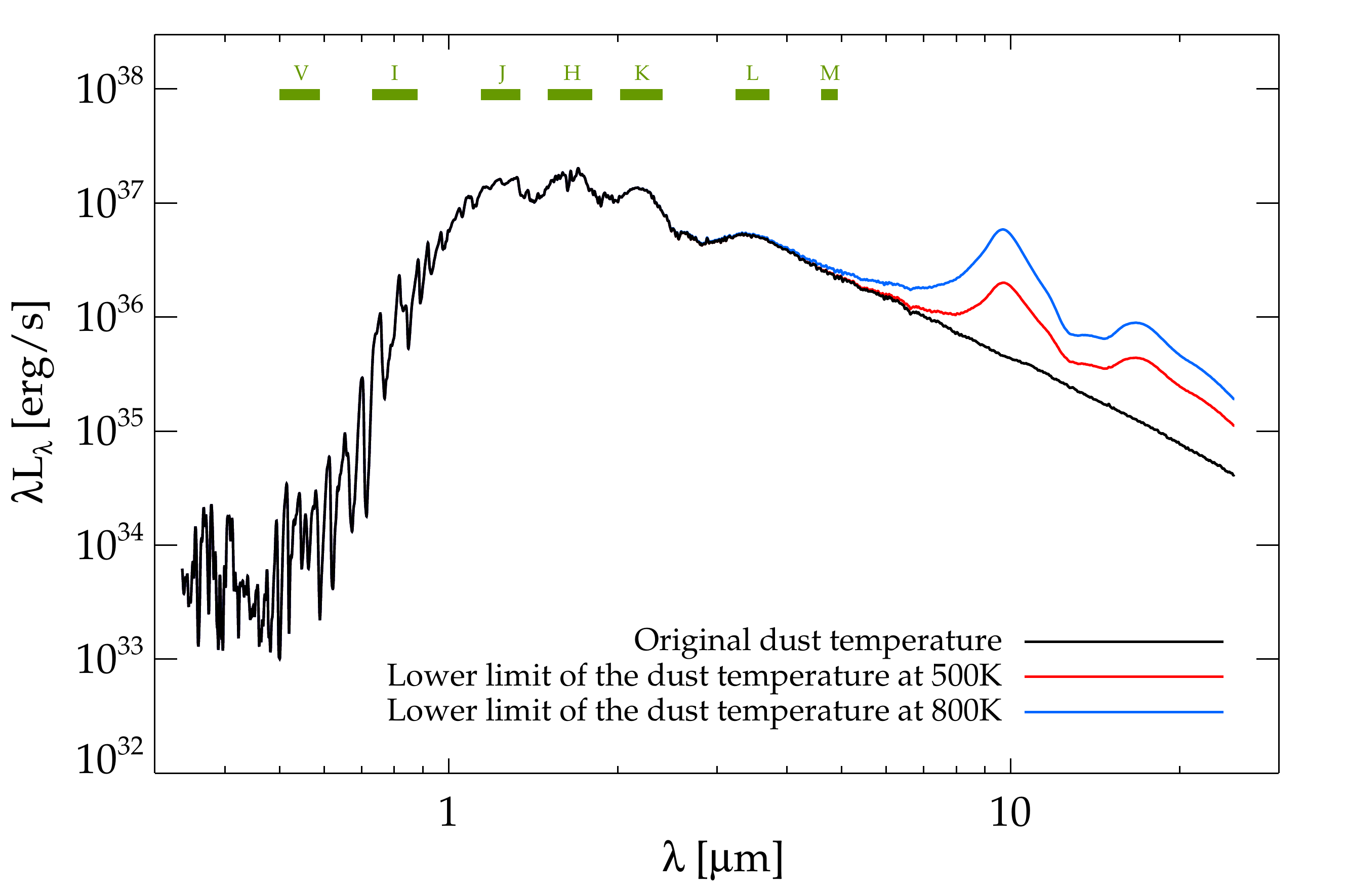}
\includegraphics[width=\linewidth]{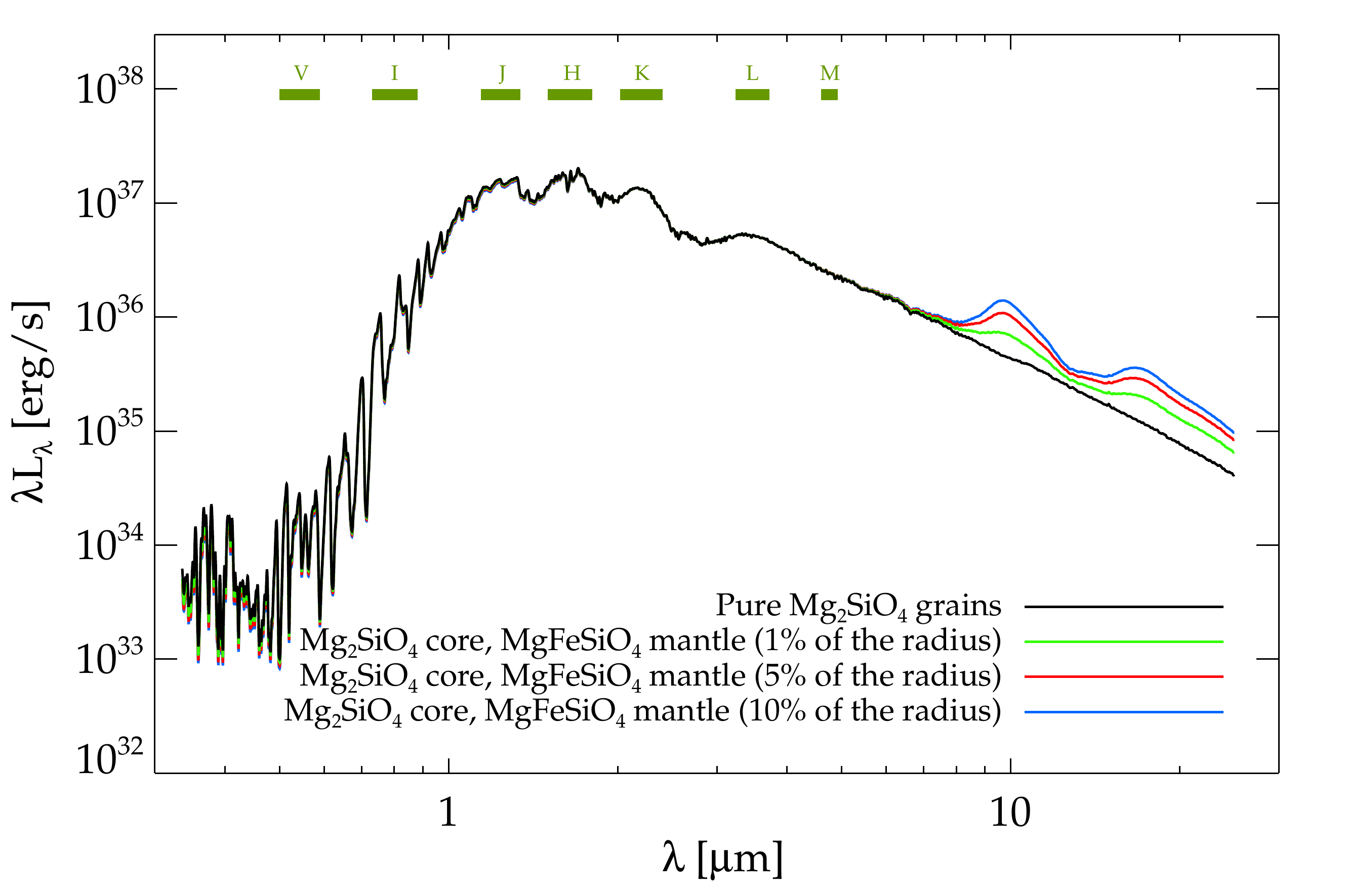}
\caption{Spectral energy distributions as a function of wavelength resulting from the dust temperatures indicated in Fig.~\ref{f_temp}. The black curve shows the original spectrum for a dynamical model with pure Mg$_2$SiO$_4$ grains (\mbox{$M=1\,\mathrm{M}_{\odot}$}, \mbox{$L=5000\,\mathrm{L}_{\odot}$} and \mbox{$T_{*}=2800\,$K}). \textit{Top panel:} The spectra produced when setting a minimum grain temperature of 800\,K (blue) and 500\,K (red) for the pure Mg$_2$SiO$_4$ grains, i.e. using the temperature structure in the top panel of Fig.~\ref{f_temp}. \textit{Bottom panel:} The spectra produced when using a core of Mg$_2$SiO$_4$ and a mantle of MgFeSiO$_4$. The green, red and blue lines show the spectra for a mantle size of 1\%, 5\% and 10\% of the radius, respectively. The corresponding temperature structures are shown in the lower panels of Fig.~\ref{f_temp}}
    \label{f_spec}
\end{figure}

To test this scenario we use a snapshot of the radial structure from a dynamical model where the wind is driven by photon scattering on Mg$_2$SiO$_4$  grains. In the \textit{a posteriori} radiative transfer calculations we set a lower limit on the grain temperature at 800\,K or 500\,K, as if the grains contained impurities that would keep them warm. The top panel of \mbox{Fig. \ref{f_temp}} shows the original temperature structure for the Mg$_2$SiO$_4$ particles, as well as the cases where a minimum dust temperature of 800\,K and 500\,K is set. The resulting spectra are shown in the top panel of \mbox{Fig. \ref{f_spec}}. It is clear from the prominent silicate features produced in the latter cases that the missing spectral features are not caused by a low abundance of silicate dust in the dynamical models, but rather a too low grain temperature. Keeping the grains warm results in silicate emission features. The warmer the grains are, the more prominent the features.

As mentioned above, a way to increase the grain temperature is to include impurities of Fe in the Mg$_2$SiO$_4$ grains. This can be simulated by considering dust particles with a core of Mg$_2$SiO$_4$ and a thin mantle of MgFeSiO$_4$, as if a small fraction of Fe had condensed onto the surface of pure magnesium silicate grains.\footnote{It is possible to introduce this modification in the \textit{a posteriori} radiative transfer since the inclusion of a small amount of other materials into these particles should not alter the overall structure of the models significantly; transparent materials will not have enough radiative cross-section to alter the energy balance and opaque materials are only thermally stable further out in the wind.}  We determine the temperature of these grains through radiative equilibrium, assuming that the addition of a thin mantle MgFeSiO$_4$ will not significantly change the structure of the dynamical model. In the \textit{a posteriori} spectral calculation we replace the temperature structure and the optical properties of the Mg$_2$SiO$_4$ grains with the corresponding properties for the core-mantle grains from the point where they are thermally stable. We assume this to be at the distance from the star where the temperature of the core-mantle grains drops below 1000\,K (approximately the temperature where magnesium-iron silicates will condense). The resulting temperature structure for three different mantle sizes (1\%, 5\% and 10\% of the radius corresponding to 3\%, 15\% and 30\% of the volume) are shown in \mbox{Fig. \ref{f_temp}} and the corresponding spectra are shown in the bottom panel of \mbox{Fig. \ref{f_spec}}. We note that dust particles with a thin mantle of MgFeSiO$_4$ may be thermally stable quite close to the stellar surface compared to pure MgFeSiO$_4$ grains. For the minimum luminosity phase of this particular model the core-mantle grains become thermally stable at $\sim 2.5\,$R$_*$, $\sim 4.5\,$R$_*$ and $\sim 5.5\,$R$_*$, respectively, if the mantle is 1\%, 5\% or 10\% of the grain radius. The overall temperature structure and, consequently, the radial distances at which dust particles are thermally stable, will vary during the pulsation cycle, and depend on the stellar parameters.

\begin{figure}
\centering
\includegraphics[width=\linewidth]{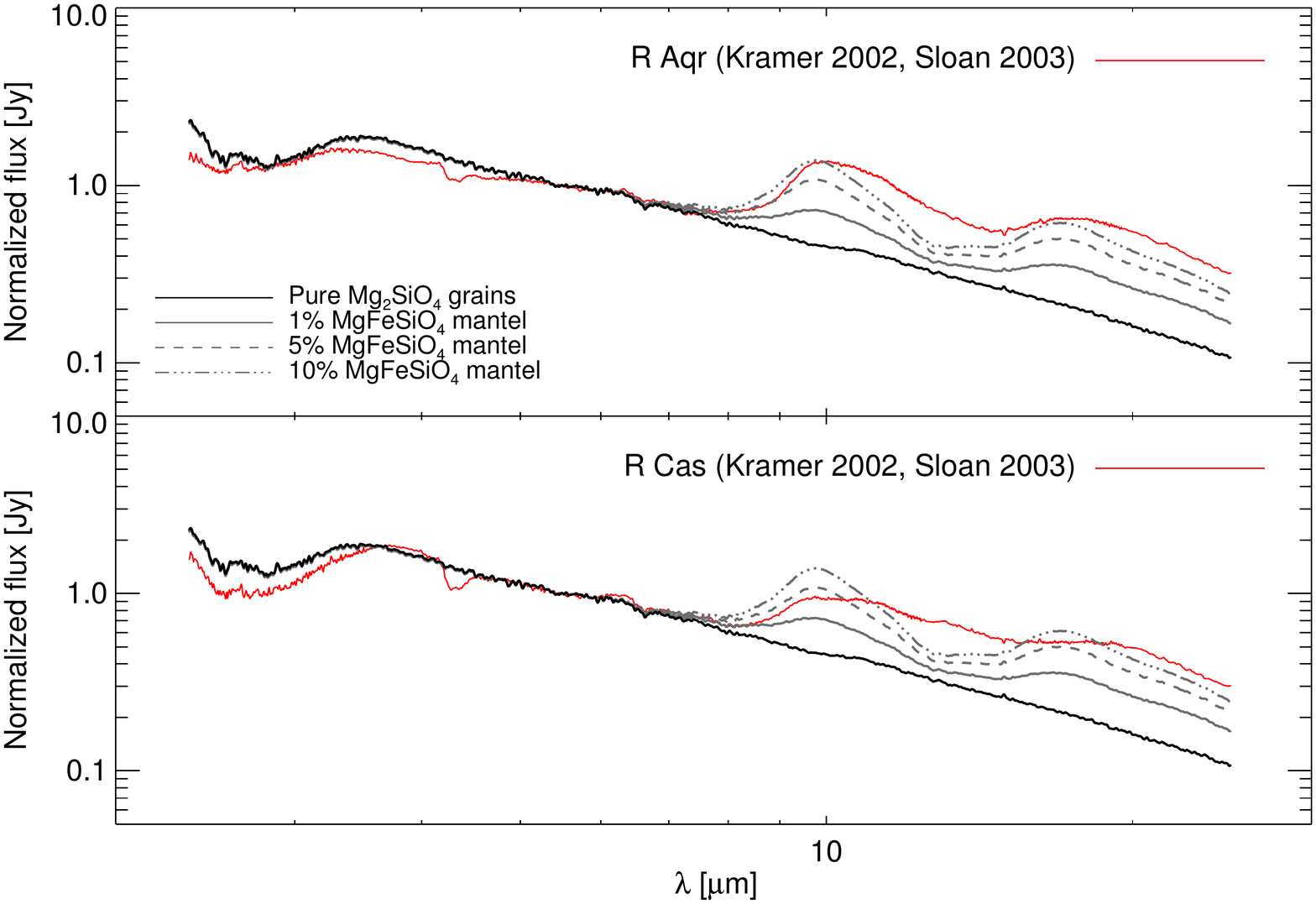}
   \caption{Spectral energy distribution as a function of wavelength. The red curves shows the observed ISO/SWS spectra of R Aqr and R Cas \citep{kram1,sloan1}. The black solid curve shows the spectra from a dynamical model using pure Mg$_2$SiO$_4$ grains and the grey curves show the spectra when assuming a core of Mg$_2$SiO$_4$ and a mantle of MgFeSiO$_4$ in the a posteriori spectral calculations. The dashed, dot-dashed and dotted lines show the spectra setting the mantle size to 1\%, 5\% and 10\% of the grain radius, respectively.}
    \label{f_spec4}
\end{figure}

To summarise the results from the second test, it seems that a thin mantle of Fe-bearing silicates, in this case consisting of MgFeSiO$_4$, can be thermally stable quite close to the stellar surface. This thin mantle of MgFeSiO$_4$ corresponds to a small Fe/Mg-ratio but is enough to heat the dust particles and produce silicate features without changing the spectra in the visual and near-IR significantly. For illustration, Fig \ref{f_spec4} shows the ISO spectra of two M-type AGB stars, R Aqr and R Cas, in addition to the original model spectrum and the model spectra with a thin mantle of MgFeSiO$_4$ on top of the Mg$_2$SiO$_4$ grains.

\section{Grain temperature and the onset of outflow}
\label{s_gtemp}
The grain temperature in our dynamical models is assumed to be determined by radiative equilibrium, but the low absorption cross-section of Mg$_2$SiO$_4$ in the wavelength region where AGB stars emit most of their flux raises the question if this assumption is realistic. \cite{gaug90} have estimated the relative importance of different processes (radiative, collisional, chemical) for changing the internal energy of dust particles consisting of amorphous carbon or astronomical silicates. They found that radiative processes dominate by many orders of magnitude over collisional processes in both cases. However, these materials have high absorption cross-sections in the visual and near-IR wavelength regions, leading to efficient radiative heating. In Appendix \ref{s_radcol} we give estimates for the radiative and collisional heating rates of Mg$_2$SiO$_4$ grains: while radiative heating is much less efficient than for the materials tested by \cite{gaug90}, it still clearly dominates over heating by collisions with the warmer gas.

\begin{figure}
\centering
\includegraphics[width=\linewidth]{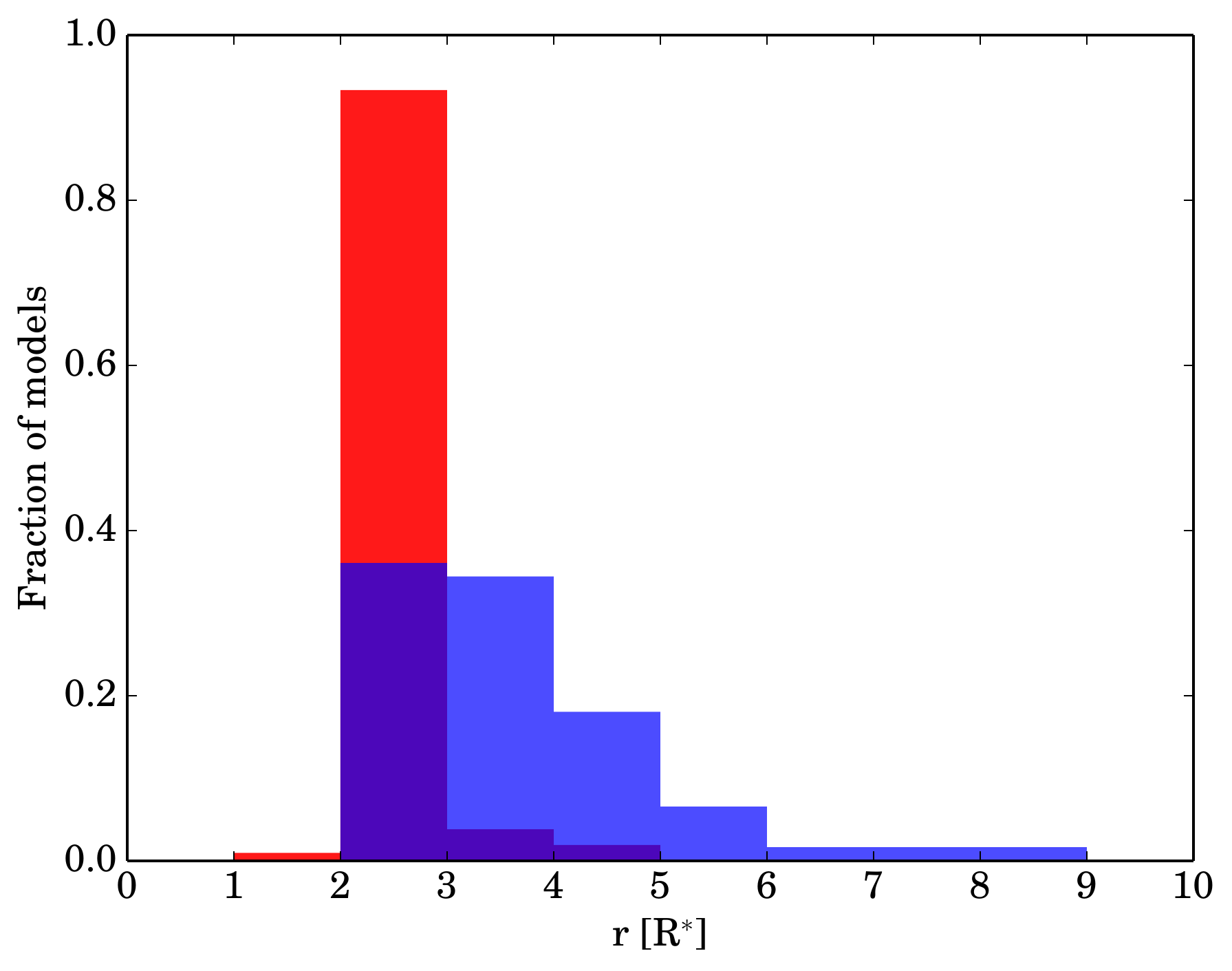}
   \caption{Position of the onset of outflow for all models developing a wind, averaged over the pulsation cycle. The red and blue areas represent models where the grain temperature is determined by radiative equilibrium and models where the grain temperature is set equal to the gas temperature, respectively. The purple area is where the two types of models overlap.}
    \label{f_wrad}
\end{figure}

Under radiative equilibrium conditions, grains of this material tend to cool rapidly in the wind (because of their optical properties, as discussed in Sect.~\ref{s_mir}), with grain temperatures generally a few hundred degrees below the gas temperature. Setting the grain temperature equal to the warmer gas temperature, instead of assuming radiative equilibrium, results in shifting the position of the dust formation zone slightly outwards. This provides us with an opportunity to explore how the radial position of the wind acceleration zone affects dynamic and photometric properties. Fig.~\ref{f_wrad} shows the position of the onset of outflow, averaged over a pulsation cycle, for the two types of models. The red area shows the results from models with a grain temperature determined by radiative equilibrium, indicating a typical distance of the onset of outflows of about $2-3\,R_*$, whereas the blue area shows the results from the models where the grain temperature is set equal to the gas temperature, indicating typical distances of about $3-6\,R_*$ (the purple area shows the overlapping region).

\begin{figure}
\centering
\includegraphics[width=\linewidth]{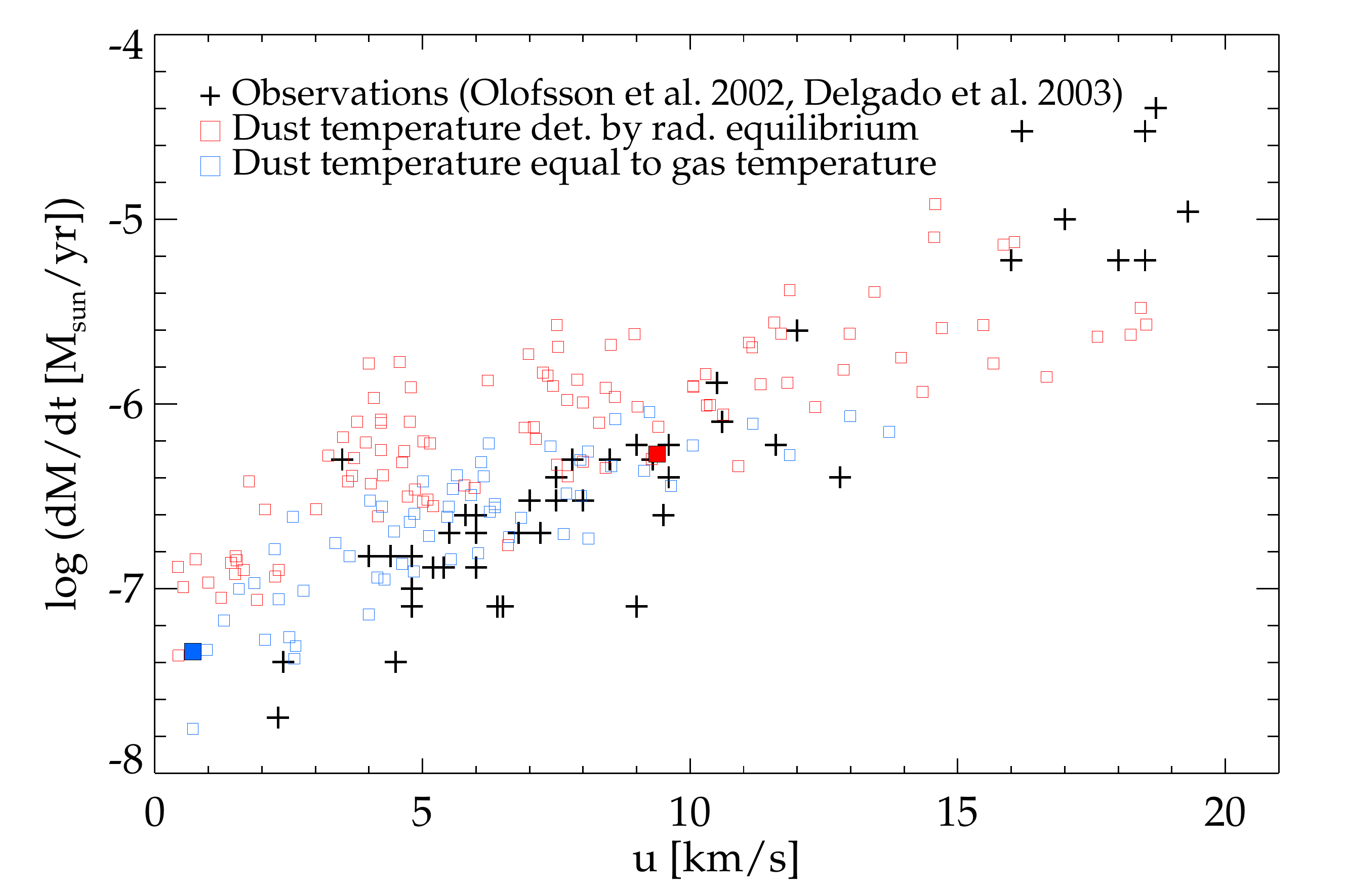}
   \caption{Observed mass-loss rates vs. wind velocities of M-type AGB stars \citep[][plus signs]{hans02,gondel03} and the corresponding properties for the dynamical models where the dust temperature is determined by radiative equilibrium (red squares) or set equal to the gas temperature (blue squares). The filled squares mark the dynamical models where a more detailed analysis is conducted.}
    \label{f_dt1}
    \quad\\
\includegraphics[width=\linewidth]{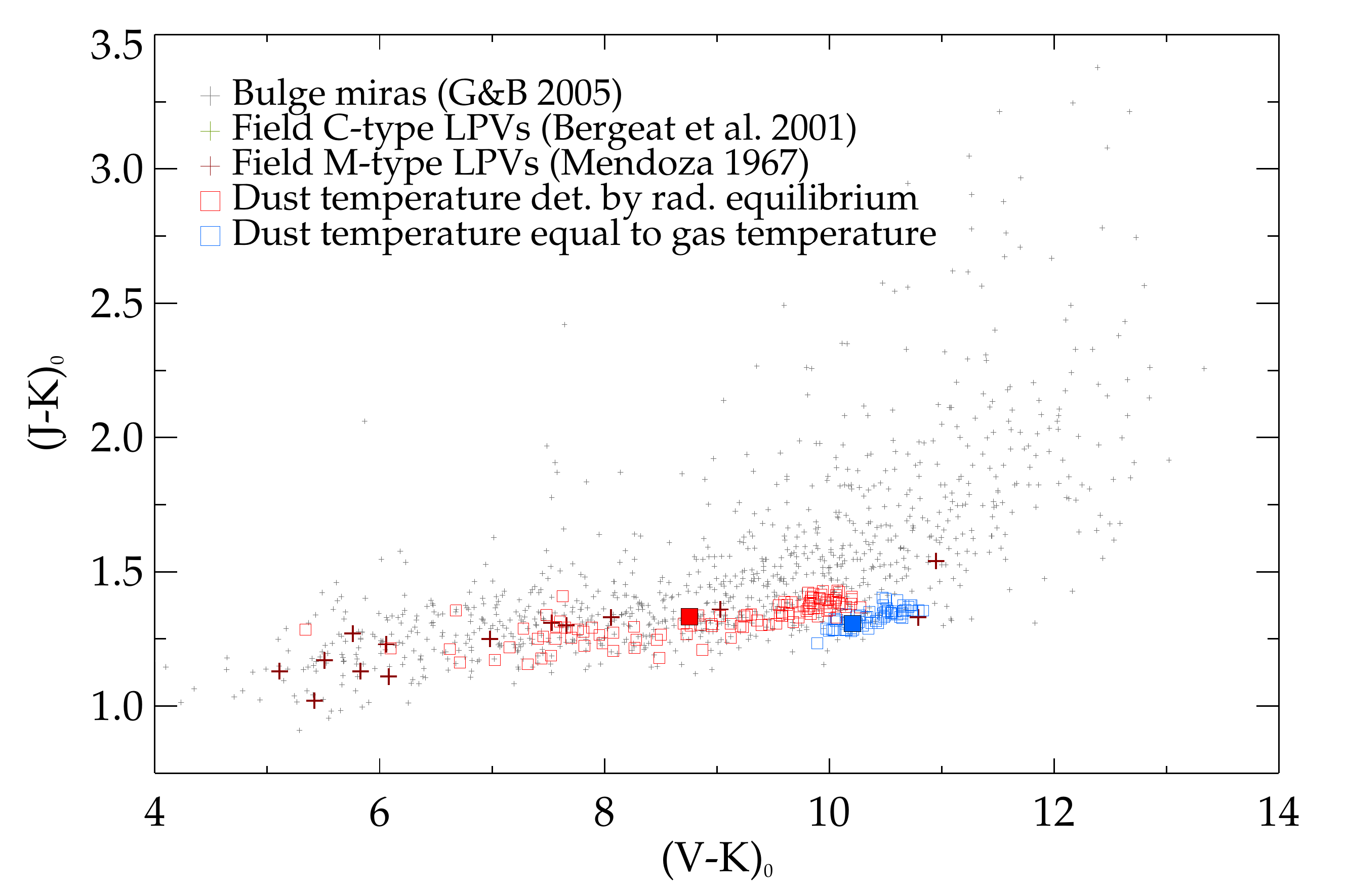}
\includegraphics[width=\linewidth]{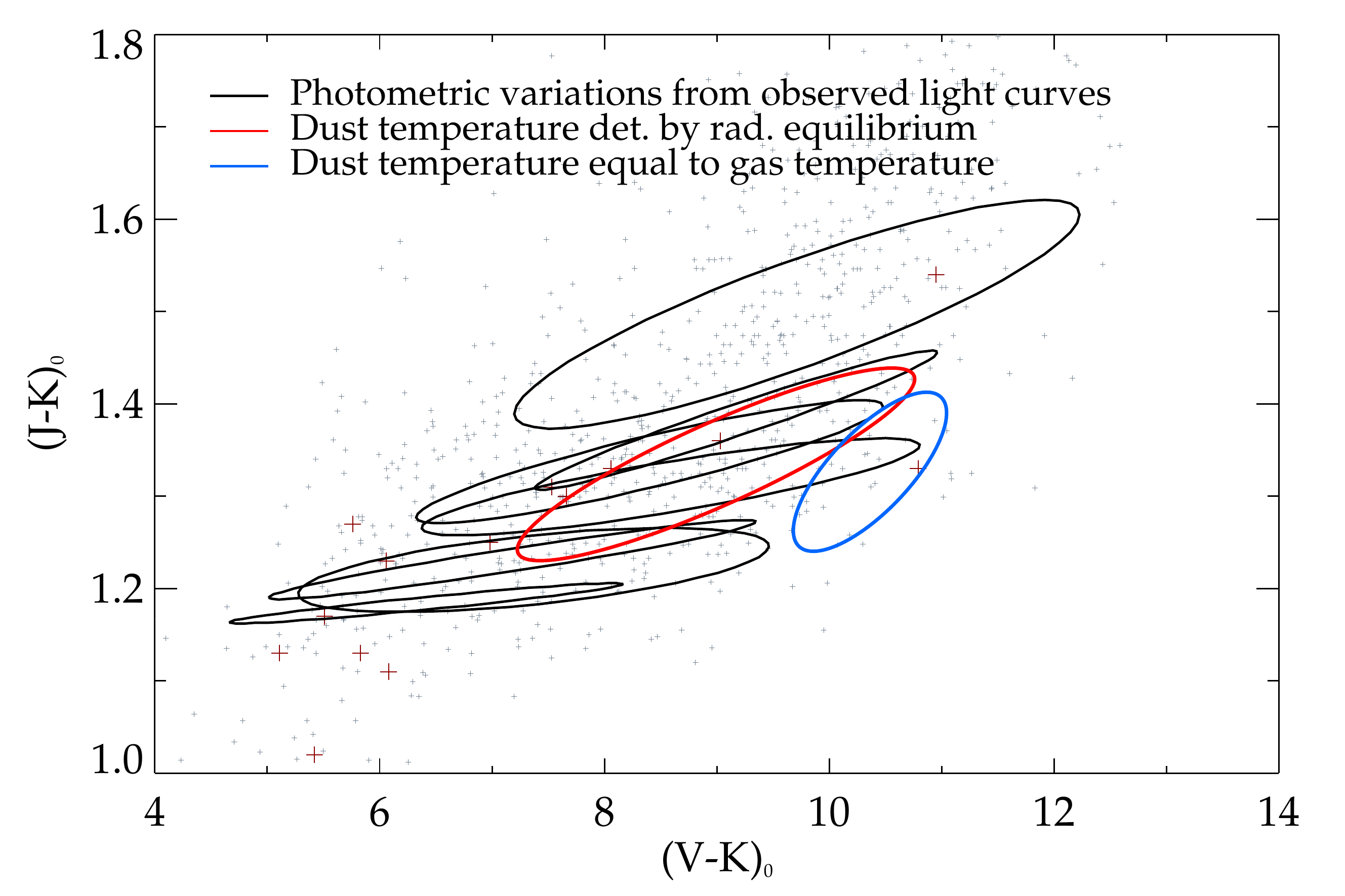}
   \caption{Observed and synthetic visual and near-IR colours. \textit{Top panel:} Synthetic phase-averaged colours from wind models where the dust temperature is determined by radiative processes (red squares) and the gas temperature (blue squares). \textit{Bottom panel:} Photometric variations derived from sine fits of observed light-curves for a sample set of M-type miras. The synthetic photometric variations for the models marked with filled squares (\mbox{$M=1\,\mathrm{M}_{\odot}$}, \mbox{$L=5000\,\mathrm{L}_{\odot}$}, \mbox{$T_{*}=2800\,$K}, $u_\mathrm{p}=4$ km/s and $\log n_{\mathrm{gr}}/n_{\mathrm{H}}=-15$) in the top panel are also plotted (solid red and blue loops). The colours are calculated from sine fits of light-curves, same as for the observational data. See \cite{bladh13} for more details concerning the sine-fit of observed light curves. For details on the observational data see Fig.~\ref{f_grid4}.}
    \label{f_dt2}
\end{figure}

This shift in the radial position of the wind acceleration zone affects the observable properties of the dynamical models. Fig.~\ref{f_dt1} shows the mass-loss rates and wind velocities of the models with a dust temperature based on radiative equilibrium (red squares) and the models where we have assumed that the dust temperature is equal to the gas temperature (blue squares). Both fit observations reasonably well, but the blue models (where the wind acceleration zone is shifted outwards to lower densities) do not reproduce the observations with high values of mass-loss rate and wind velocity, while the stars with low mass loss and slow wind velocities seem to be better reproduced.

In contrast to the dynamical properties, photometry provides strong constraints on the position of the wind acceleration zone. The top panel of Fig.~\ref{f_dt2} shows the average colours $(J-K)$ and $(V-K)$ during a pulsation cycle for the two types of models. The photometry of the models where the grain temperature is set equal to the gas temperature (blue squares) is very similar to the photometry for pulsating models without wind (see Fig.~4 in \cite{bladh13}), whereas the photometry of the models based on radiative equilibrium (red squares) are bluer in $(V-K)$. An in-depth comparison is given in the lower panel of Fig.~\ref{f_dt2}, showing the photometric variations during a pulsation cycle. The red and blue loops are from two dynamical models where the dust temperature is set in different ways but which otherwise have the same model parameters. The black loops show observed photometric variations from a set of very regular mira stars (for a more detailed description of the observational data see \cite{bladh13}). The large variations in $(V-K)$ seen in the observed photometry are not reproduced in the models with a grain temperature set equal to the gas temperature.

It should be mentioned here again that the effects on the photometric colours are not due to dust opacities. Instead the visual and near-IR photometry of the models gives direct diagnostics of the gas, i.e. the structure and dynamics of the atmosphere. As mentioned in Sect.\ref{s_nir}, the large variations observed in $(V-K)$ are a consequence of a change in the abundance of TiO during the pulsation cycle. The position of the onset of outflow affects the overall density and temperature structure of the gas and, consequently, the molecular abundances, by radially expanding the atmosphere outwards. If the wind acceleration zone is shifted to distances where this effect no longer influences the structure of the inner atmosphere, then the resulting photometry will resemble that of a pulsating atmosphere without a stellar wind. It seems that the observed variations in $(V-K)$ do not only reveal that the wind-driving dust species in M-type AGB stars have to be quite transparent in the visual and near-IR, but they also provide an upper limit for the distance of the onset of outflow. We note that this upper limit probably also holds if a different mechanism than radiation pressure on dust accelerates the wind, since the diagnostic presented here is based on gas properties.

\section{Summary and conclusions}
\label{s_sum}

In this paper we present the first extensive set of time-dependent models for atmospheres and winds of M-type AGB stars. The stellar winds in these models are driven by photon scattering on Mg$_2$SiO$_4$ dust. Grains of this material can form close enough to the stellar surface, they consist of abundant materials and, if they grow to sizes comparable to the wavelength of the stellar flux maximum, can provide sufficient radiative acceleration to drive outflows. 

The current study includes 139 solar-mass models, using three different luminosities ($5000\,$L$_{\odot}$, $7000\,$L$_{\odot}$ and $10000\,$L$_{\odot}$) and effective temperatures ranging from 2600\,K to 3200\,K (see Table~\ref{t_grid}). The input parameters describing stellar pulsations and seed particle abundance are also varied for each combination of stellar parameters, resulting in different atmospheric structures and wind properties. The models produce outflows for a wide range of stellar parameters (see Fig.~\ref{f_ifwind}), as well as mass-loss rates, wind velocities, and visual and near-IR photometry in good agreement with observations (see Fig.~\ref{f_dyn2} and Fig.~\ref{f_grid2}). This demonstrates that winds driven by photon scattering on Mg$_2$SiO$_4$ grains is a viable scenario for the mass loss of M-type AGB stars. However, the current models do not show the characteristic silicate features at \mbox{10 and 18 $\mu$m}. This is not caused by a low abundance of silicate dust in the dynamical models, but rather a too rapidly falling temperature of Mg$_2$SiO$_4$ grains. Adding a thin mantle of MgFeSiO$_4$ to these grains, at distances where such a mantle can be thermally stable, increases the grain temperature and results in spectra with pronounced silicate features. 

The visual and near-IR spectra provide constraints on the driving mechanism of the outflows. Stellar winds driven by photon scattering will produce very different spectra than winds driven by true absorption, as there is much less circumstellar reddening by the dust component in the former case. This is confirmed by observed and synthetic colours of M-type AGB stars; the visual and near-IR fluxes are characterised by large variations in $(V-K)$ and small variations in $(J-K)$ during the pulsation cycle (see Fig.~\ref{f_grid4}). The variations in $(V-K)$ are caused by changes in the molecular abundances (mostly TiO) and not by changes in the dust opacities \citep[see Fig.~\ref{f_spectio} and][]{bladh13}. From this we can infer that the envelopes of M-type AGB stars, and, consequently, the dust grains, are quite transparent in the visual and near-IR wavelengths, otherwise the spectral changes would not be dominated by molecular features. The results regarding the mid-IR features mentioned above, however, indicate that the grains have to include a small amount of impurities (dirty silicates). Preliminary tests demonstrate that including a small amount of Fe in magnesium silicate grains is sufficient to solve the problem of grain temperatures and that the visual and near-IR photometry will not be affected significantly. Future models should include a more complete treatment of dust formation.

The models presented in this paper are an important step towards closing a long standing gap in stellar evolution theory. For the first time, it is possible to compute both mass-loss rates and the corresponding spectra from the same time-dependent atmosphere and wind models of M-type AGB stars. This is essential input for models of stellar evolution and population synthesis. However, before the model results can be generally applied in these contexts it is necessary to compute a bigger grid with better coverage of mass, luminosity, effective temperature and other fundamental parameters.

\begin{acknowledgements}
The computations were performed on resources provided by the Swedish National Infrastructure for Computing (SNIC) at UPPMAX. This work has been funded by the Swedish Research Council (\textit{Vetenskapsr\aa det}). The authors also acknowledge the support from the
{\em project STARKEY} funded by the ERC Consolidator Grant, G.A. n.~615604.
\end{acknowledgements}

\bibliographystyle{aa}
\citeindextrue
\bibliography{references}

\begin{appendix} 

%\section{Heating processes of dust}
\section{Heating of Mg$_2$SiO$_4$ grains}
\label{s_radcol}
The very low absorption cross-section of Mg$_2$SiO$_4$ grains in the wavelength region where AGB stars emit most of their flux raises the question if grain heating is dominated by radiation or collision with the warmer gas. Estimates of the processes that determine the internal energy of a dust particle, and consequently the grain temperature, have previously been done by \cite{gaug90} for amorphous carbon and astronomical silicates. Both materials have high absorption cross-sections in the near-IR and, consequently, radiative processes dominate over collisional energy exchange. The mid-IR optical properties of Mg$_2$SiO$_4$ grains (relevant for radiative cooling, see Sect.~\ref{s_mir}) are comparable to those of astronomical silicates. However, the near-IR absorption coefficients, which determine the radiative heating, are several orders of magnitudes lower than those of Fe-bearing silicates. Therefore we need to examine if radiative heating still dominates over heating by collisions with the warmer gas in the case of Mg$_2$SiO$_4$ grains.\footnote{\cite{gaug90} considered a third process, i.e. energy change due to association or disassociation of one monomer. They found this to be negligible compared to collisional processes, and similar reasoning holds for the current case.}

We start by estimating the energy exchange occurring when a dust particle of size $a$ and temperature $T_{\mathrm{d}}$ collides with a gas particle of temperature $T_{\mathrm{g}}$,
 \begin{eqnarray*}
\Delta E_{\mathrm{coll}} &=& \frac{k}{4\pi}(T_{\mathrm{g}}-T_{\mathrm{d}}),
\end{eqnarray*}
(for details see \cite{gaug90}). If we multiply the energy transfer during one collision $\Delta E_{\mathrm{coll}}$ with the number of collisions per second $\tau^{-1}_{\mathrm{coll}}$, we can estimate the rate of change in internal energy for a dust particle as a result of collisions with the surrounding gas,
\begin{eqnarray*}
L_{\mathrm{coll}} &=& \frac{1}{\tau_{\mathrm{coll}}}\Delta E_{\mathrm{coll}}= \pi a^2\,v_{\mathrm{th}}\,n_{\mathrm{H}}\,\beta\,\Delta E_{\mathrm{coll}}\,.
\end{eqnarray*} 
Here $v_{\mathrm{th}}$ is the average relative velocity between a dust particle and a hydrogen atom which, given the difference in mass between the two particles, can be approximated with the average thermal velocity of the hydrogen atom.\footnote{The hydrogen particles in the dust formation zone will probably mostly be in the form of H$_2$, but we ignore this extra complication in the following order of magnitude estimates. This means that the following numbers will overestimate the collisional heating somewhat, erring on the safe side.} The particle density of hydrogen is denoted $n_{\mathrm{H}}$ and $\beta$ is the probability that a collision will result in an energy transfer of $\Delta E_{\mathrm{coll}}$. The term $\tau^{-1}_{\mathrm{coll}}$ is a measure of how many gas particles per second collide with the cross-section $\pi a^2$ of a dust particle and transfer energy. If we assume that the particle density of hydrogen is given by $n_{\mathrm{H}}=\rho/1.4m_{\mathrm{H}}$ we get the following expression for the change in energy due to collisions
\begin{eqnarray*}
L_{\mathrm{coll}} &=& \pi a^2\,v_{\mathrm{th}}\,\beta\,\frac{\rho}{1.4m_{\mathrm{H}}}\,\frac{k}{4\pi}(T_{\mathrm{g}}-T_{\mathrm{d}})\,.
\end{eqnarray*} 
We set the grain temperature to \mbox{$T_{\mathrm{d}}=1000\,$K} and the gas temperature to \mbox{$T_{\mathrm{g}}=1500\,$K}, both typical values in the dust formation zone. Furthermore, we assume a grain size of \mbox{$a=10^{-6}\,\mathrm{cm}$} and set \mbox{$\beta=1$} (corresponding to an upper limit). The average thermal velocity for this gas temperature is \mbox{$v_{\mathrm{th}}\approx 5$\,km/s}. According to detailed dynamical models, a typical gas density is about \mbox{$\rho\approx 10^{-14}\,\mathrm{g/cm}^3$} in the dust formation zone.\footnote{This agrees well with observed mass-loss rates and wind velocities for M-type AGB stars. Using conservation of mass, \mbox{$\rho=\dot{M}/4\pi R^2v$}, and assuming a typical wind velocity of \mbox{$10$\,km/s} and mass-loss rate of \mbox{$10^{-6}\,\mathrm{M}_{\odot}/\mathrm{yr}$} results in a density of about \mbox{$\rho\approx 10^{-14}\,\mathrm{g/cm}^3$} in the dust formation zone.} Consequently, the energy gained per second due to collisions of a dust particle with the surrounding gas is estimated to
\begin{eqnarray*}
L_{\mathrm{coll}} &\approx& 4\cdot 10^{-11}\quad\mathrm{[erg/s]}\,.
\end{eqnarray*} 
Next we estimate the radiative heating rate of a dust particle. The rate of change in internal energy of a grain due to the absorption of photons from the ambient radiation field can be expressed as
 \begin{eqnarray*}
L_{\mathrm{abs}} &=& 4\pi\,\pi a^2 \!\int_{0}^{\infty} \!Q(\lambda,a)\,J_{\lambda}\,\mathrm{d}\lambda,
\end{eqnarray*}
where $a$ is the radius of the dust particle, $Q(a,\lambda)$ is the absorption efficiency and $J_{\lambda}$ is the mean intensity of the incident radiation field. In the case of \mbox{M-type} AGB stars photometric observations indicate that the stellar radiation is not significantly reddened by the dust material in the circumstellar envelope. We can therefore consider the radiation field as optically thin and approximate it by a geometrically diluted stellar radiation field, here represented by a Planck function. In the small particle limit the optical properties of a dust material $Q^{\prime}(\lambda)=Q(\lambda,a)/a$ can be approximated with a power-law function, where $D$\,[cm$^{-1}$] and $p$ are material-dependent properties,
\begin{eqnarray*}
Q^{\prime}(\lambda)&\approx& D\cdot \left(\frac{\lambda}{\lambda_0}\right)^{-p}\,.
\end{eqnarray*}
Consequently, the rate of change in internal energy due to absorption of photons for a dust particle in the small particle limit is given by
\begin{eqnarray*}
L_{\mathrm{abs}}&=&4\pi^2 a^3\int_{0}^{\infty} \!\frac{Q(\lambda)}{a}\,J_{\lambda}\,\mathrm{d}\lambda\\
&=&4\pi^2 a^3\int_{0}^{\infty} \!Q^{\prime}(\lambda)\,W(r)B_{\lambda}(T_{\mathrm{*}})\,\mathrm{d}\lambda\\
&=& W(r)\,D_1\,a^3\,\frac{8\pi^2 k^4}{c^2h^3}\left(\frac{\lambda_0k}{ch}\right)^{p}T_{\mathrm{*}}^{4+p}\int_0^{\infty}\frac{x^{3+p}}{e^x-1}dx\\
&=& W(r)\,D_1\,a^3\,\frac{8\pi^2 k^4}{c^2h^3}\left(\frac{\lambda_0k}{ch}\right)^{p}T_{\mathrm{*}}^{4+p}\Gamma (4+p)\zeta (4+p)
\end{eqnarray*}
where  $x=hc/k\lambda T$. Here $D$ and $p$ are fitted in the wavelength region where the star emits most of its flux, which is approximately at $1\,\mu$m for a star with an effective temperature \mbox{$T_*=2800\,$K}. Using the optical data for Mg$_2$SiO$_4$ given by \cite{jag03} we derive material constants \mbox{$D\approx 3\,\mathrm{cm}^{-1}$} and \mbox{$p\approx -0.9$} for $\lambda_0=1\,\mu$m. Since Mg$_2$SiO$_4$ grains condense around \mbox{$r\approx 2\,R_*$} the geometric dilution factor is approximately \mbox{$W(r)=\frac{1}{2}\left[1-\sqrt{1-(R_*/r)^2}\right]\approx 0.07$}. Setting $a=10^{-6}\,$cm, and \mbox{$\Gamma (4+p)\zeta (4+p)\approx 2.6$} we estimate the rate of change in internal energy for a dust particle due to absorption of photons to be
\begin{eqnarray*}
L_{\mathrm{abs}}&\approx& 2\cdot 10^{-8}\,\mathrm{erg/s}\,.
\end{eqnarray*}
It should be noted that this number is probably a lower limit since the power law fit of $Q'$ used here tends to underestimate the contribution to grain heating from the Mid-IR wavelength region. There the lower stellar flux may be compensated by higher grain opacity around $10\,\mu$m (see Fig.~\ref{f_flux}). 

Given these simple estimates we can safely assume that radiative heating dominates over collisional heating for magnesium silicates, even though the absorption cross-section of these grains is very low in the wavelength range where M-type AGB stars emit the most of their flux. These estimates are for quite small grains ($0.01\,\mu$m) but should hold for even larger grains, given the functional dependence of each process on grain size.

\end{appendix}

\end{document}